\documentclass[aps,prb,reprint,superscriptaddress,amsmath,amssymb]{revtex4-1}

\usepackage{graphicx}
\usepackage{float}
\def\ket#1{|#1\rangle }
\begin{document}




\title{
Nonsymmorphic Dirac semimetal and carrier dynamics in doped spin-orbit-coupled Mott insulator Sr$_2$IrO$_4$
}

\author{J. W. Han }
\thanks{These two authors contributed equally to this work.}
\affiliation{Department of Physics and Photon Science, School of Physics and Chemistry, Gwangju Institute of Science and Technology (GIST), Gwangju 61005, Republic of Korea}

\author{Sun-Woo Kim}
\thanks{These two authors contributed equally to this work.}
\affiliation{Department of Physics, Hanyang University, Seoul 04763, Republic of Korea}

\author{W. S. Kyung}
\affiliation{Advanced Light Source, Lawrence Berkeley National Laboratory, California 94720, USA}
\affiliation{Center for Correlated Electron Systems, Institute for Basic Science (IBS), Seoul 08826, Republic of Korea}
\affiliation{Department of Physics and Astronomy, Seoul National University (SNU), Seoul 08826, Republic of Korea }

\author{C. Kim}
\affiliation{Center for Correlated Electron Systems, Institute for Basic Science (IBS), Seoul 08826, Republic of Korea}
\affiliation{Department of Physics and Astronomy, Seoul National University (SNU), Seoul 08826, Republic of Korea }

\author{G. Cao}
\affiliation{Department of Physics, University of Colorado, Boulder, Colorado 80309, USA  }

\author{X. Chen}
\affiliation{Department of Materials, University of California, Santa Barbara, California 93106, USA }

\author{S. D. Wilson}
\affiliation{Department of Materials, University of California, Santa Barbara, California 93106, USA }

\author{Sangmo Cheon}
\email{sangmocheon@hanyang.ac.kr}
\affiliation{Department of Physics, Hanyang University, Seoul 04763, Republic of Korea}

\author{J. S. Lee}
\email{jsl@gist.ac.kr}
\affiliation{Department of Physics and Photon Science, School of Physics and Chemistry, Gwangju Institute of Science and Technology (GIST), Gwangju 61005, Republic of Korea}

\begin{abstract}
A Dirac fermion emerges as a result of interplay between symmetry and topology in condensed matter. 
Current research moves towards investigating the Dirac fermions in the presence of many-body effects in correlated system. 
Here, we demonstrate the emergence of correlation-induced symmetry-protected Dirac semimetal state in the lightly-doped spin-orbit-coupled Mott insulator Sr$_2$IrO$_4$. 
We find that the nonsymmorphic crystalline symmetry stabilizes a Dirac line-node semimetal and that the correlation-induced symmetry-breaking electronic order further leads to a phase transition from the Dirac line-node to a Dirac point-node semimetal. 
The latter state is experimentally confirmed by angle-resolved photoemission spectroscopy and terahertz spectroscopy on Sr$_2$(Ir,Tb)O$_4$ and (Sr,La)$_2$IrO$_4$.
Remarkably, the electrodynamics of the massless Dirac carriers is governed by the extremely small scattering rate of about 6 cm$^{-1}$ even at room temperature, which is iconic behavior of relativistic quasiparticles. 
Temperature-dependent changes in electrodynamic parameters are also consistently explained based on the Dirac point-node semimetal state.
\end{abstract}

\maketitle


Dirac semimetal (DSM) \cite{armitage_weyl_2018} is a new quantum state of matter protected by the interplay of symmetry and topology, and exhibits the electrodynamics of relativistic quasiparticles in a condensed matter system. 
Such a novel state of quantum matter is described by the massless Dirac equation and extensive investigations \cite{kotov_electron-electron_2012,elias_dirac_2011,tang_role_2018,ye_massive_2018,yin_giant_2018,fujioka_strong-correlation_2019} have been made about how the Dirac physics incorporates electron correlation and spin-orbit coupling (SOC). 
When the electron correlation and/or SOC is strong, a Dirac fermion acquires its velocity renormalization \cite{elias_dirac_2011,tang_role_2018} or mass \cite{ye_massive_2018,yin_giant_2018}. 
For instance, a Dirac fermion in graphene near the charge-neutrality point undergoes logarithmic velocity renormalization by long-range Coulomb interaction \cite{elias_dirac_2011,tang_role_2018} and a Dirac fermion in the correlated kagome magnet becomes massive in the presence of strong SOC \cite{ye_massive_2018,yin_giant_2018}. 
The interactions involved determine the fate of the Dirac fermions. 
It is thus interesting to investigate how the interplay among electron correlation, SOC, and crystalline symmetries affects Dirac fermions. 
So far, most studies \cite{elias_dirac_2011,tang_role_2018,ye_massive_2018,yin_giant_2018,fujioka_strong-correlation_2019} of Dirac fermions in a correlated system have been limited to pure (undoped) materials which are inherent metals. 
Unlike pure systems, a doped correlated system having, for example, the remnant correlations endemic to the parent Mott state \cite{imada_metal-insulator_1998,chakravarty_hidden_2001,lee_doping_2006,keimer_quantum_2015,fradkin_colloquium:_2015}, is particularly promising for seeing the interplay of various kinds of interactions owing to the emergence of a plethora of correlation-induced symmetry-breaking orders \cite{chakravarty_hidden_2001,lee_doping_2006,keimer_quantum_2015,fradkin_colloquium:_2015}. 
In this respect, a doped correlated system hosting Dirac fermions could provide a new window for studying the relatively unexplored realm of correlated Dirac fermions. 

Recently, iridates have attracted considerable attention due to the novel correlated and topological phases arising from a delicate combination of electron correlation and SOC \cite{witczak-krempa_correlated_2014,rau_spin-orbit_2016}. 
Among them, a spin-orbit-coupled Mott insulator Sr$_2$IrO$_4$ \cite{kim_novel_2008,kim_phase-sensitive_2009} is of particular interest because of its striking similarity to cuprate phenomenology under doping
—such as Fermi arcs \cite{kim_fermi_2014,kim_observation_2016}, a d-wave pseudogap \cite{kim_fermi_2014,kim_observation_2016,de_la_torre_collapse_2015}, and various symmetry-breaking orders \cite{zhao_evidence_2016,battisti_universality_2017,zhou_correlation_2017,chen_unidirectional_2018}. 
Here, by combining theoretical calculations with angle-resolved photoemission spectroscopy (ARPES) and terahertz (THz) spectroscopy on Sr$_2$(Ir,Tb)O$_4$ and (Sr,La)$_2$IrO$_4$, 
we demonstrate that the essential Dirac semimetal protected by nonsymmorphic crystalline symmetry emerges in the lightly-doped spin-orbit-coupled Mott insulator Sr$_2$IrO$_4$ even in the presence of strong SOC and electron correlation. 
Remarkably, a correlation-induced symmetry-breaking order leads to a phase transition from a Dirac line node (DLN) to a Dirac point node (DPN). 
The correlation effect is also manifest in temperature-dependent Dirac carrier dynamics with an extremely small scattering rate, enabling the distinction between DLN and DPN states.

\begin{figure*}[t]
\includegraphics{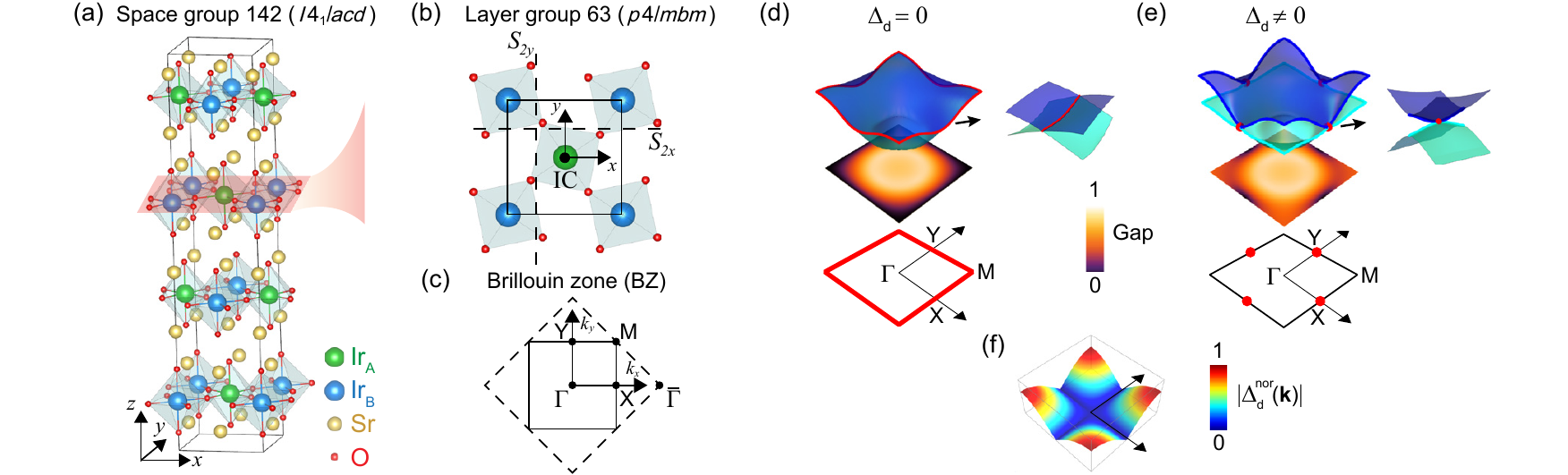}
\caption{(a),(b) Crystal structure of layered and single-layer Sr$_2$IrO$_4$.
(c) BZ for $\sqrt{2}\times\sqrt{2}$ (1$\times$1) unit cell drawn by solid (dashed) line.
(d),(e) Calculated 3D band structures of single-layer Sr$_2$IrO$_4$ without and with d-wave order.
Below each 3D band structure, the 2D contour plot shows the normalized band gap. 
Red lines and points indicate fourfold-degeneracy.
(f) Normalized d-wave gap function $\Delta_\text{d}^{\text{nor}} (\mathbf{k}) = \sin(k_x/2) \sin(k_y/2)$. 
}
\end{figure*}

Sr$_2$IrO$_4$ is a layered compound, where each layer consists of Ir$_\text{A}$ and Ir$_\text{B}$ sublattices surrounded by oxygen octahedra that are rotated around the $z$-axis in a staggered fashion forming a $\sqrt{2}\times\sqrt{2}$ unit cell [Figs. 1(a) and 1(b)]. 
Owing to the weak interlayer interaction, Sr$_2$IrO$_4$ (space group $I4_1/acd$) is a quasi-two-dimensional (quasi-2D) system whose crystalline symmetries are captured by the layer group $p4/mbm$ \cite{supple}. 
The ground state of Sr$_2$IrO$_4$ is an antiferromagnetic Mott insulating state \cite{kim_phase-sensitive_2009}. 
Upon Tb or La doping \cite{wang_decoupling_2015,chen_influence_2015}, the antiferromagnetism is strongly suppressed and a metallic state appears even at low doping concentration. 
For our 3$\%$ Tb-doped Sr$_2$IrO$_4$ system, Sr$_2$(Ir$_{0.97}$Tb$_{0.03}$)O$_4$, it is reported that Tb$^{4+}$ (4$f^7$) ions substitute for Ir$^{4+}$ ions while keeping the crystal symmetry of the parent compound \cite{wang_decoupling_2015}. 
The isovalent Tb$^{4+}$ dopants do not introduce additional charge carriers and completely suppress the N{\'e}el temperature, inducing a paramagnetic metallic state with time-reversal symmetry. 

To study the symmetry-protected DSM in Sr$_2$(Ir$_{0.97}$Tb$_{0.03}$)O$_4$, we first perform a symmetry analysis for the layer group $p4/mbm$ with time-reversal symmetry. 
This layer group contains symmetry elements of inversion $P$, mirror $M_z$, and two orthogonal nonsymmorphic screw rotations $S_{2x}\equiv\{C_{2x}|\frac{1}{2}00\}$ and $S_{2y}\equiv\{C_{2y}|0\frac{1}{2}0\}$
---i.e., a rotation $C_{2x,2y}$ around the $x$-or $y$-axis followed by a translation of half a lattice vector along that $x$-or $y$-axis [Fig. 1(b)]. 
Note that each rotation axis of nonsymmorphic symmetries $S_{2x}$ and $S_{2y}$ does not intersect an inversion center (IC). 
This kind of nonsymmorphic symmetry is called off-centered nonsymmorphic symmetry \cite{wieder_spin-orbit_2016,yang_topological_2017}. 
Let us see the role of the single off-centered screw rotation $S_{2x}$.
The commutation relation between $P$ and $S_{2x}$ is given by
$PS_{2x}=e^{-ik_x+ik_y}S_{2x}P$, 
leading to $\{P,S_{2x}\}=0$
at X and Y points in the Brillouin zone (BZ). 
Combined with inversion $P^2=1$ and time-reversal $\Theta^2=-1$, this symmetry algebra guarantees fourfold-degenerate Dirac point nodes (DPNs) at X and Y points \cite{wieder_spin-orbit_2016,supple}. 
With $M_z$ satisfying $\{M_z,S_{2x}\}=0$,
the DPN at the Y point is further spanned as a Dirac line node (DLN) along the $k_y=\pi$ line \cite{supple}. 
Similarly, $S_{2y}$ with $P\Theta$ and $M_z$ symmetries generates DLN along the $k_x=\pi$ line. 
Therefore, the multiple symmetries of $S_{2x}$, $S_{2y}$, $P\Theta$, and $M_z$ protect the DLN along the whole BZ boundary, which is confirmed by the calculated band structure [Fig. 1(d)].

Next, we consider the electron correlation effect in lightly Tb-doped Mott system. 
The most prominent correlated electron phenomenon in lightly-doped Mott systems is the emergence of a pseudogap as ubiquitously observed in lightly-doped cuprates \cite{lee_doping_2006,keimer_quantum_2015,fradkin_colloquium:_2015}.
Indeed, for lightly electron-doped Sr$_2$IrO$_4$, a d-wave pseudogap was experimentally observed \cite{kim_fermi_2014,kim_observation_2016,de_la_torre_collapse_2015}
and was plausibly explained by symmetry-breaking electronic order
---a so-called d-wave spin-orbit density wave (d-SODW) \cite{zhou_correlation_2017}---in the mean-field level. 
In Tb-doped system, we also have experimental evidence for a d-wave pseudogap in ARPES experiment, as we will see. Thus, we introduce d-SODW order with a d-wave form factor 
$\Delta_\text{d} (\mathbf{k}) = -4 \Delta_\text{d} \sin(k_x/2) \sin(k_y/2)$
 [Fig. 1(f)] to consider the electron correlation-induced pseudogap \cite{supple}.

When the correlation-induced d-wave order is present ($\Delta_\text{d} \neq 0$), 
$S_{2x}$ and $S_{2y}$ symmetries are broken, while $M_z$ and $P\Theta$ symmetries are preserved \cite{supple}.
This leads to the lifting of the fourfold degeneracy of DLN. However, from the nature of a d-wave form factor [Fig. 1(f)], 
there are symmetry-invariant lines $k_x=0$ and $k_y=0$ where the d-wave order vanishes
---that is, $S_{2x}$ and $S_{2y}$ symmetries are still maintained. 
Hence, at X and Y points, the previous symmetry algebra for $S_{2x}$ and $S_{2y}$ holds, 
which guarantees fourfold-degenerate states, i.e., $\ket{\Psi}, P \ket{\Psi}, \Theta \ket{\Psi}$, and  $P \Theta \ket{\Psi}$.
Consequently, the electron correlation-induced d-wave order lifts the fourfold degeneracies along the whole BZ boundary, 
except for the points located at the symmetry-invariant lines, 
leading to the phase transition from DLN to DPN [Fig. 1(e)] ~\cite{comment_NS_DSM}. 
The relativistic dispersion relation $E=pc$ is established along only the 1D direction for DLN, 
whereas a complete 2D relativistic dispersion relation is realized in DPN. 
Due to this dimensional discrepancy, differences in physical observables are expected between DLN and DPN.

\begin{figure}[t]
\includegraphics{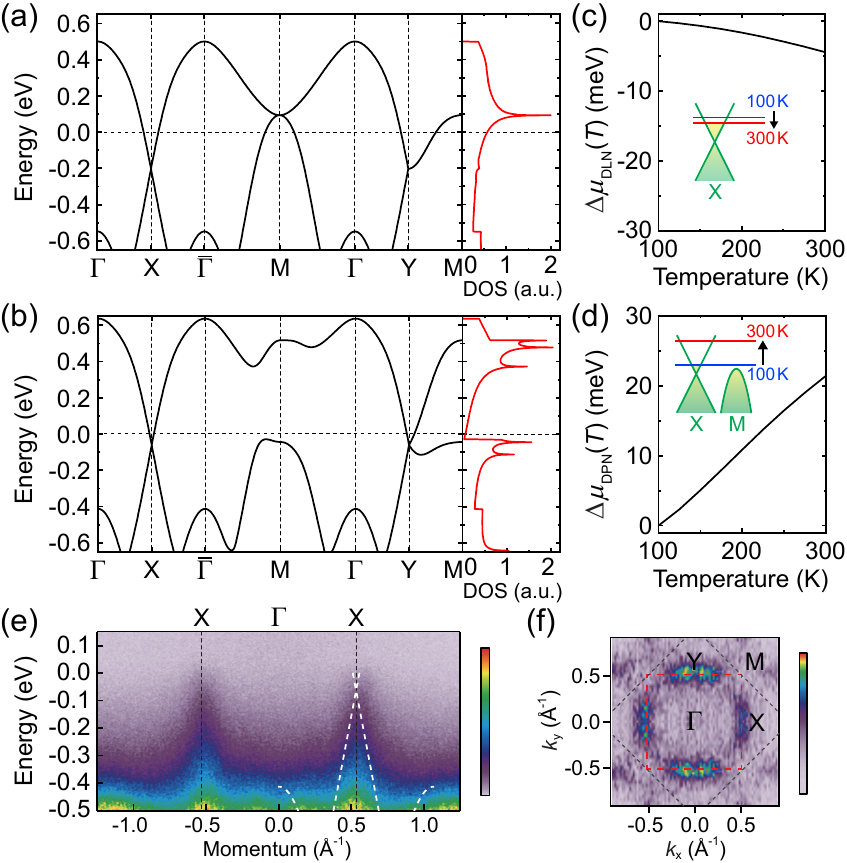}
\caption{
(a),(b) Calculated band structures and DOS for (a) DLN and (b) DPN states. 
(c),(d) Calculated $T$-dependent chemical potentials $\mu (T)$ of (c) DLN and (d) DPN states.
(e) ARPES spectra of Sr$_2$(Ir$_{0.97}$Tb$_{0.03}$)O$_4$ at 100 K along X$-\Gamma-$X line, which are overlaid with the band structure of the DPN state (white dashed lines). Note that the measured Fermi velocity ($\approx$ 1/1000 c) agrees well with the theoretical value. (f) Constant binding energy $k_{\text x}-k_{\text y}$ map at $E=E_{\text F}$. 
}
\end{figure}

For a quantitative comparison with experiment, 
we calculate the electronic structures of DLN and DPN using a realistic 2D tight-binding model including SOC with hopping parameters derived from density-functional theory calculations \cite{zhou_correlation_2017}.
We treat electron correlation effects within a five-orbital mean-field Hubbard model \cite{supple}.
Let us first see the electronic structure without d-wave order. 
In Fig. 2(a), the band structure shows a dispersive DLN along the $\overline{\text{YM}}$ line. 
At the M point, two parabolic bands meet, leading to a van Hove singularity (vHs) near energy $E=0.1$ eV, as shown in the density of states (DOS). 
Meanwhile, the correlation-induced d-wave order lifts the degeneracy of DLN along the BZ boundary, forming anisotropic DPNs at X and Y points [Figs. 1(e) and 2(b)]. 
At the M point, the two parabolic bands become gapped with an energy gap of $\sim0.4$ eV, 
generating several vHs as shown in DOS [Fig. 2(b)]. 
The correlation-induced vHs near the Fermi level ($E\sim-0.04$ eV), in particular, gives rise to an unconventional temperature- ($T$-) dependent chemical potential.
Calculated chemical potentials $\mu_{\text{DLN}} (T)$ and $\mu_{\text{DPN}} (T)$ of DLN and DPN states show the opposite $T$-dependence [Figs. 2(c) and 2(d)].
In this case, $\mu_{\text{DLN}} (T)$ decreases with increasing $T$.
On the other hand, $\mu_{\text{DPN}} (T)$ is proportional to $T$
---different from $\mu(T)$ of graphene [Fig. S15(a) in Supplemental Material \cite{supple}]---
because the correlation-induced heavy hole band at the M point acts as an effective carrier reservoir supplying electrons to the Dirac band. 
Note that these chemical potentials of DLN and DPN are essential ingredients for determining differences in $T$-dependent Dirac carrier dynamics.

\begin{figure}[b]
\includegraphics{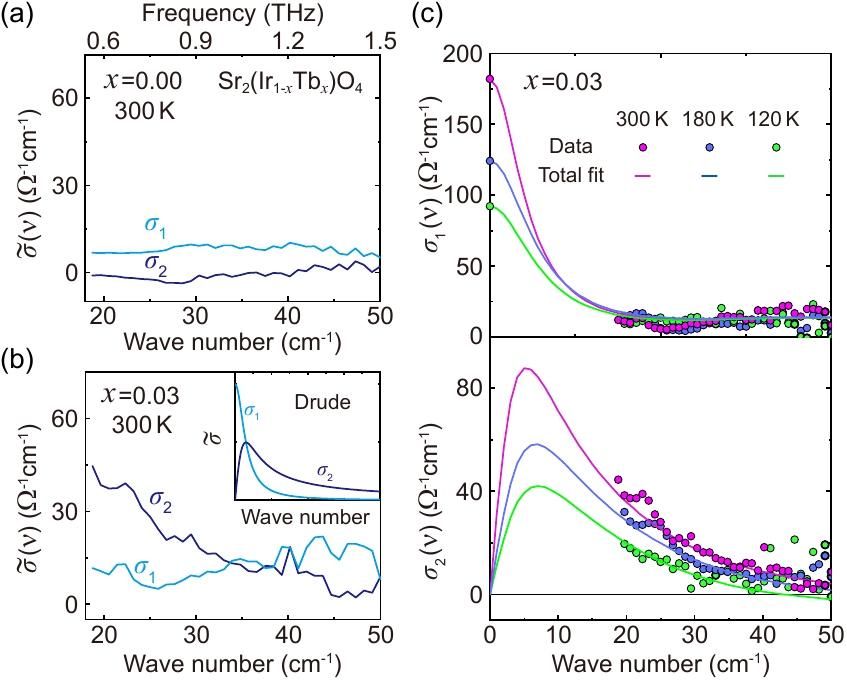}
\caption{
(a),(b) Real ($\sigma_1 (\nu)$) and imaginary ($\sigma_2 (\nu)$) parts of $\tilde{\sigma}$ obtained at 300 K for (a) $x=0.00$ and (b) $x=0.03$.
The inset in (b) shows the typical Drude response. 
(c) $\sigma_1 (\nu)$ and $\sigma_2 (\nu)$ of $x=0.03$ at 120, 180, and 300 K.
Symbols at zero wave number indicate the DC conductivity values obtained from the transport measurement \cite{wang_decoupling_2015}.
Solid lines are Drude-Lorentz fitting results where each contribution of Drude and Lorentz response is presented in Supplemental Material \cite{supple}. 
}
\end{figure}

Before addressing Dirac carrier dynamics, 
we identify the electronic structure of Sr$_2$(Ir$_{0.97}$Tb$_{0.03}$)O$_4$ using ARPES.
As shown in Fig. 2(e), 
Dirac-like linear dispersion is observed up to the Fermi level with a Dirac point at the X point with the binding energy $\sim 50$ meV (see also Supplemental Material \cite{supple}). 
The 2D ARPES map in Fig. 2(f) shows a clear d-wave pseudogap feature, 
namely, a gap at the M point and nodal regions around X and Y points, 
which is simliar with that observed in electron-doped Sr$_2$IrO$_4$ \cite{kim_fermi_2014,kim_observation_2016,de_la_torre_collapse_2015}.
These ARPES results are consistent with the calculated band structure with d-wave order in Fig. 2(b). 
Therefore, the ARPES data support the emergence of DPN rather than DLN~\cite{comment_DLN}.

Next, we study Dirac carrier dynamics and further distinguish the electronic states of Tb-doped Sr$_2$IrO$_4$ based on THz spectroscopy. 
To figure out changes of carrier dynamics upon Tb doping, 
we first study the Mott insulating state of the parent Sr$_2$IrO$_4$ at 300 K [Fig. 3(a)]. 
The real part $\sigma_1 (\nu)$ of the optical conductivity $\tilde{\sigma} (\nu)$ shows a featureless spectral response from 20 to 50 cm$^{-1}$ with a successful connection to the DC conductivity $\sigma_{\text{DC}} \sim 10 ~ \Omega^{-1}$cm$^{-1}$ \cite{wang_decoupling_2015}. Similarly, the imaginary part $\sigma_2 (\nu)$ shows a flat spectrum. 
These spectral features are consistent with the Mott insulating state of the parent compound \cite{moon_dimensionality-controlled_2008}. 
For Sr$_2$(Ir$_{0.97}$Tb$_{0.03}$)O$_4$ [Fig. 3(b)], $\sigma_1 (\nu)$ rises slightly with increasing frequency.
In contrast to the parent Sr$_2$IrO$_4$, a significant increase of $\sigma_1 (\nu)$ at $<$ 20 cm$^{-1}$ is expected because $\sigma_{\text{DC}} = 180~ \Omega^{-1}$cm$^{-1}$ [Fig. 3(c)]. 
Consistently, $\sigma_2 (\nu)$ increases with decreasing frequency. 
These spectral observations are reminiscent of the high frequency tail of the Drude response; wide spectral features of the Drude peak are displayed in the inset of Fig. 3(b).
It should be noted that the scattering rate defined in the Drude model corresponds to the half-width of $\sigma_1$ and the extremal point frequency of $\sigma_2$, or equivalently the crossing point of $\sigma_1$ and $\sigma_2$. 
In these respects, the results in Fig. 3(b) clearly indicate that the free carrier scattering rate is given to be far less than 20 cm$^{-1}$. 

We employ the Drude-Lorentz model to analyze $\tilde{\sigma} (\nu)$ 
and extract two important electrodynamics parameters, the scattering rate $\gamma$ and Drude weight $D$.
We verify the validity of the model detailed in Supplemental Material \cite{supple}. 
Note that the fitting has been performed by satisfying $\sigma_1 (\nu)$, $\sigma_2 (\nu)$, and $\sigma_{\text{DC}}$ simultaneously.  
At 300 K, we obtain $\gamma=5.8$ cm$^{-1}$ and $D= 62500$ cm$^{-2}$
[Fig. 3(c)]. 
It is worth emphasizing that the observed $\gamma$ is extremely small even at room temperature, 
and it is one of key fingerprints of Dirac fermions; 
$\gamma$ is significantly reduced due to the large Fermi velocity 
and the chiral nature of massless Dirac fermions \cite{castro_neto_electronic_2009,das_sarma_electronic_2011}. 
Similar small values of $\gamma$ have been reported for several DSMs \cite{bolotin_temperature-dependent_2008,chen_optical_2015,crassee_nonuniform_2018}. 
For a 2D DSM, the suspended graphene \cite{bolotin_temperature-dependent_2008} exhibits $\gamma \approx 20$ cm$^{-1}$ at 40 K. 
For 3D DSM, ZrTe$_5$ \cite{chen_optical_2015} and Cd$_3$As$_2$ \cite{crassee_nonuniform_2018} show $\gamma \approx 13$ and 10 cm$^{-1}$ at 200 and 300 K, respectively.

As discussed before, differences in $T$-dependent carrier dynamics between DLN and DPN are expected, 
and hence THz optical spectra are examined in detail with varying $T$. 
In Fig. 3(c), we compare $\tilde{\sigma}(\nu)$ obtained at 120, 180, and 300 K.
$\sigma_1 (\nu)$ does not show a noticeable $T$-dependent spectral change. 
However, $\sigma_2 (\nu)$ becomes larger with increasing $T$, thereby showing the systematic increase of the spectral weight. 
We fit all measured $\tilde{\sigma} (\nu)$ \cite{supple} and obtain $T$-dependent scattering rate $\gamma (T)$ and Drude weight $D(T)$ as displayed in Figs. 4(a) and 4(b), respectively. 
$\gamma (T)$ decreases with increasing $T$ and a very small $\gamma (T)$ is observed over a wide $T$ range,
consistent with the fact that a symmetry-protected Dirac semi-metallic state is robust against thermal effect. 
$D(T)$ increases with increasing $T$. 
Thus, the combination of $\gamma(T)$ and $D(T)$ leads to the rise of $\sigma_1 (\nu \rightarrow 0)$ with increasing $T$, 
in good agreement with the transport data shown in Fig. 4(c). 

\begin{figure}
\includegraphics{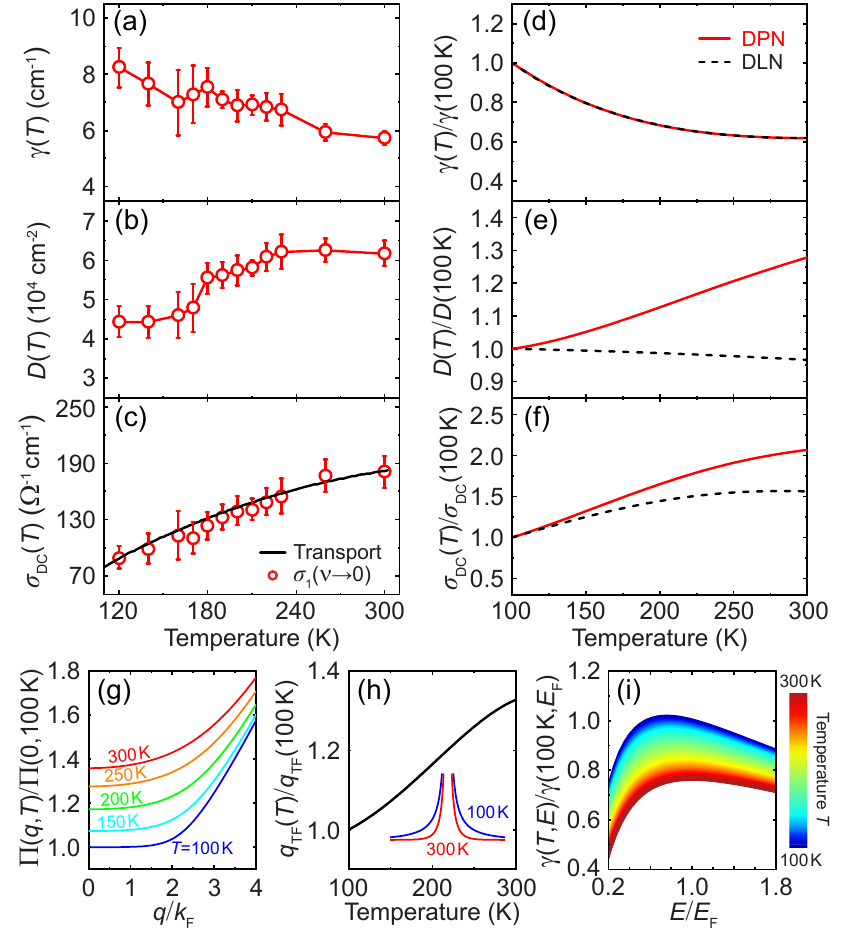}
\caption{(a),(b) Experimentally obtained scattering rate $\gamma (T)$ and Drude weight $D(T)$. 
(c) DC conductivity $\sigma_{\text{DC}} (T)$ estimated by Drude analysis (symbol) and directly acquired by transport measurement (line). 
(d)-(f) Theoretically obtained $\gamma (T)$, $D(T)$, and $\sigma_{\text{DC}} (T)$ \cite{supple}.
(g)-(i) Calculated polarizability function $\Pi(q,T)$, 
 Thomas-Fermi wave vector $q_{\text{TF}}(T)$,
and RPA scattering rate $\gamma(T,E)$ for DPN state \cite{supple}.
}
\end{figure}

To understand the THz data measured, 
we calculate the $T$-dependent optical conductivity using the Boltzmann transport equation \cite{supple}. For scattering mechanisms in Dirac systems (e.g., graphene), 
one can consider both phonon and charged impurity scatterings  \cite{das_sarma_electronic_2011}. 
However, decreasing $\gamma(T)$ with increasing $T$ [Fig. 4(a)] cannot be explained by phonons which inevitably result in increasing $\gamma(T)$ with increasing $T$. 
Thus, we focus on the charged impurity scattering and consider two types of charged impurities depending on the nature of their Coulomb potentials: short-range and long-range charged impurities. 
By combining these two charged impurities, we construct a phenomenological model for the scattering rate $\gamma_{\text{model}} (T,E)$ as described in Supplemental Material \cite{supple}. 
Then, using $\gamma_{\text{model}} (T,E)$ as input, we calculate $\gamma(T)$ for DLN and DPN [Fig. 4(d)] such that the calculated $\gamma(T)$ for both DLN and DPN are well-matched with the experimental values in Fig. 4(a). 
With the same $\gamma(T)$ values, the calculated $\sigma_{\text{DC}} (T)$ for DLN and DPN show qualitatively similar tendency with $T$ [Fig. 4(f)], which means that $\sigma_{\text{DC}} (T)$ alone is insufficient to distinguish between DLN and DPN. 
Instead, as shown in Fig. 4(e), the calculated $D(T)$ for DLN and DPN show the opposite $T$-dependence; $D(T)$ for DPN exhibits a consistent $T$-dependence with the experimental result [Fig. 4(b)]. 
This is mainly due to the different $T$-dependent chemical potentials between DLN and DPN [Figs. 2(c) and 2(d)]. 
Upon increasing $T$, $\mu_{\text{DPN}} (T)$ increases but $\mu_{\text{DLN}} (T)$ decreases, 
which gives rise to the increase or decrease of the Dirac carrier density, and equivalently, of the $D(T)$. 
Thus, from the $D(T)$ result, we conclude that the THz experiment supports the emergence of DPN rather than DLN, consistently with the ARPES experiment.

To study the microscopic scattering mechanism in the DPN state of Tb-doped system, 
we scrutinize the finite-temperature screening effect arising from Dirac fermions around charged impurities using random phase approximation (RPA). 
Within RPA, the static dielectric function $\epsilon(q,T)$ is given by $\epsilon(q,T) = 1 + v_c (q) \Pi(q,T)$ 
where $v_c(q) = 2\pi e^2 / \kappa q$ with the effective dielectric constant $\kappa$ 
and $\Pi (q,T)$ is the finitie-temperature polarizability function \cite{hwang_screening-induced_2009,supple}.
In Fig. 4(g), the calculated  $\Pi(q,T)$ increases monotonically with temperature at all $q$ owing to the thermal excitation of Dirac fermions from the valence band to the conduction band. 
Thus, the Coulomb potential is more screened with increasing $T$, 
which is characterized by the Thomas-Fermi screening wave vector
$q_{\text{TF}}(T) \equiv \lim_{q\rightarrow0} q v_c \Pi(q,T) $ [Fig. 4(h)]. 
With this $q_{\text{TF}} (T)$, we calculate the $T$-and $E$-dependent scattering rate $\gamma(T,E)$ [Fig. 4(i)];
$\gamma(T,E)$ decreases with increasing $T$ at all $E$ because the scattering cross-section is reduced by enhanced screening of the Coulomb potential,
which explains the experimental data in Fig. 4(a). 
This insulating behavior of $\gamma(T,E)$ is analogous to the high-temperature behavior of $\gamma_{\text{gr}}(T,E)$ in high-mobility graphene \cite{hwang_screening-induced_2009,bolotin_temperature-dependent_2008,supple}.  
Note that the calculated electrodynamic parameters using $\gamma(T,E)$ are consistent with both experimental and theoretical data in Fig. 4 \cite{supple}.

To generalize the emergence of correlated DSM in the lightly-doped Mott insulator Sr$_2$IrO$_4$, 
we also study a 5.5\% La-doped system, (Sr$_{0.945}$La$_{0.055}$)$_{2}$IrO$_{4}$. 
Because it has the same crystal structure and magnetic state as the Tb-doped system, 
the same symmetry analysis with d-wave order can be applied. 
Indeed, Dirac dispersion and a d-wave pseudogap were observed by ARPES for (Sr$_{0.95}$La$_{0.05}$)$_{2}$IrO$_{4}$ \cite{de_la_torre_collapse_2015}, suggesting that it has a DPN ground state. 
From the THz experiment on (Sr$_{0.945}$La$_{0.055}$)$_{2}$IrO$_{4}$, we obtain 
an extremely small scattering rate of $\sim 4$ cm$^{-1}$ at room $T$ \cite{supple}, which is the direct manifestation of Dirac carrier dynamics.

In summary, we demonstrated the universal emergence of nonsymmorphic Dirac semimetal with point-node dispersion in the lightly-doped spin-orbit-coupled Mott insulator Sr$_2$IrO$_4$. 
The interplay between nonsymmorphic symmetry and correlation-induced symmetry-breaking electronic order is essential in realizing Dirac point-node semimetal. 
Such Dirac semimetal allows the relativistic electrodynamics governed by the extremely small scattering rate even at room temperature, 
thus providing an intriguing opportunity to explore new emergent and collective phenomena of correlated Dirac materials.

This work was supported by the Science Research Center and the Basic Science Research Program through the National Research Foundation of Korea (NRF) funded by the Ministry of Science, 
ICT \& Future Planning (Nos. 2015R1A5A1009962, 2015R1A1A1A05001560). 
S.-W.K. and S.C. were supported by NRF through Basic Science Research Programs (no. NRF-2018R1C1B6007607), the research fund of Hanyang University (HY-2017), and the POSCO Science Fellowship of POSCO TJ Park Foundation. 
S.D.W. and X.C. acknowledge support from NSF Award No. DMR-1505549. 

\begingroup
\renewcommand{\addcontentsline}[3]{}

\endgroup

\clearpage

\pagebreak

\onecolumngrid
\widetext
\renewcommand{\Vec}[1]{\mbox{\boldmath$#1$}}
\def\infinity{\infty}
\def\t#1{\textrm{#1}}
\def\ket#1{|#1\rangle }
\def\bra#1{\langle #1 |}
\def\n{\nonumber \\ }
\def\tensor{\otimes}
\newcommand\abs[1]{\left|#1\right|}
\newcommand{\overbar}[1]{\mkern 1.5mu\overline{\mkern-1.5mu#1\mkern-1.5mu}\mkern 1.5mu}
\newcommand{\red}[1]{{\textcolor{red}{#1}}}
\newcommand{\blue}[1]{{\textcolor{blue}{#1}}}
\newcommand{\magenta}[1]{{\textcolor{magenta}{#1}}}

\renewcommand{\theequation}{S\arabic{equation}}
\renewcommand{\thefigure}{S\arabic{figure}}
\renewcommand{\bibnumfmt}[1]{[#1]}
\renewcommand{\citenumfont}[1]{#1}
\setlength{\belowcaptionskip}{-5pt}

\renewcommand{\thesection}{\normalsize \arabic{section}}
\renewcommand{\thesubsection}{\thesection.\arabic{subsection}}
\renewcommand{\thesubsubsection}{\thesubsection.\arabic{subsubsection}}

\begin{center}
\textbf{
Supplemental Material: Nonsymmorphic Dirac semimetal and carrier dynamics in doped spin-orbit-coupled Mott insulator Sr$_2$IrO$_4$
}
\end{center}

\tableofcontents
\addtocontents{toc}{\protect\setcounter{tocdepth}{5}}
\newpage 

\section{\normalsize S\lowercase{ymmetry analysis}}

\subsection{ Space and layer groups of Tb-doped Sr$_2$IrO$_4$}

\begin{figure}[h]
\includegraphics[width=\textwidth]{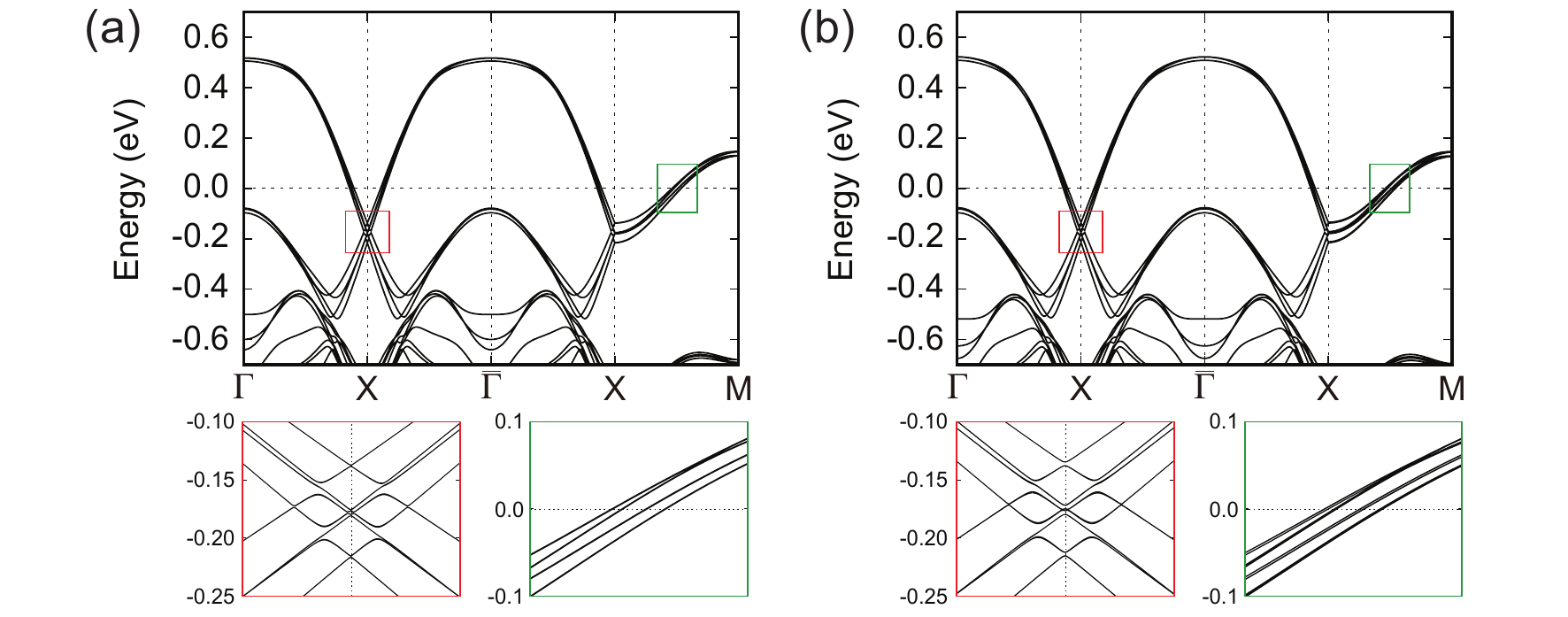} 
\caption{ 
\label{fig:DFT_band}
(a),(b) Calculated DFT band structures of paramagnetic Sr$_2$IrO$_4$ using reported experimental lattice constants for (a) I4$_1$$/acd$ structure \cite{crawford_structural_1994} and (b) I4$_1$$/a$ structure \cite{torchinsky_structural_2015}. Here, the band structures are obtained by VASP code \cite{kresse_ab_1993,kresse_efficiency_1996} with 12$\times$12$\times$2 k-points. We used PBE functional \cite{perdew_generalized_1996} including spin-orbit coupling with on-site Hubbard $U=1.6$ eV \cite{liu_electron_2016}.} 
\end{figure}

The space group of Sr$_2$IrO$_4$ was known as I4$_1$$/acd$ (no. 142) by neutron powder diffraction measurements \cite{crawford_structural_1994}
and the physical interpretations of various theoretical and experimental studies were performed based on this space group.
Meanwhile, recent experiments on single crystal have reported a small structure change, proposing that the correct space group of Sr$_2$IrO$_4$ is an I4$_1$$/a$ (no. 88) \cite{ye_magnetic_2013,ye_structure_2015,torchinsky_structural_2015}.
The change of space group from I4$_1$$/acd$ to I4$_1$$/a$ is attributed to the inequivalent tetragonal distortion between Ir$_\text{A}$ and Ir$_\text{B}$ sublattices. 
However, the observed Ir$_\text{A}$-O$_{apical}$ (Ir$_\text{A}$-O$_{basal}$) and Ir$_\text{B}$-O$_{apical}$ (Ir$_\text{B}$-O$_{basal}$) bond lengths only differ by 2.6$\times$$10^{-4}$ (1.5$\times$$10^{-3}$) \AA, which is expected to lead small change on band structure. 
Indeed, in our first-principles density-functional theory (DFT) calculations for the paramagnetic Sr$_2$IrO$_4$  show that such small structural differences barely affect the band structure as shown in Fig. \ref{fig:DFT_band}. 
In Fig. \ref{fig:DFT_band}(a), for I4$_1$$/acd$ space group, we find that the Dirac point at the X point is gapped within $\sim$ 1 meV arising from an inherent weak interlayer interaction in the layered Sr$_2$IrO$_4$. 
For I4$_1$$/a$ space group, we find that the gap at the Dirac point X further increases to $\sim$ 3 meV [Fig. \ref{fig:DFT_band}(b)]. 
It is noteworthy that the linearity of Dirac bands is still maintained, which governs the Dirac carrier dynamics at Fermi level.
Thus, the physical properties of Sr$_2$IrO$_4$  can be described using space group I4$_1$$/acd$ within a tolerance of $\sim$ 2 meV and
the difference between two space groups can be considered as a small perturbation if needed.

Moreover, one can focus on the single layer because it dictates most bulk properties owing to the weak interlayer interaction. 
For the single layer analysis, a layer group which is a subperiodic group of the space group can be introduced. 
In terms of the layer group, the layer group p$4/mbm$ (no. 63) captures the full crystalline symmetries of single layer Sr$_2$IrO$_4$ of the space group I4$_1$$/acd$.

\subsection{Symmetry analysis with and without d-wave order}
\begin{figure}[h]
\includegraphics[width=\textwidth]{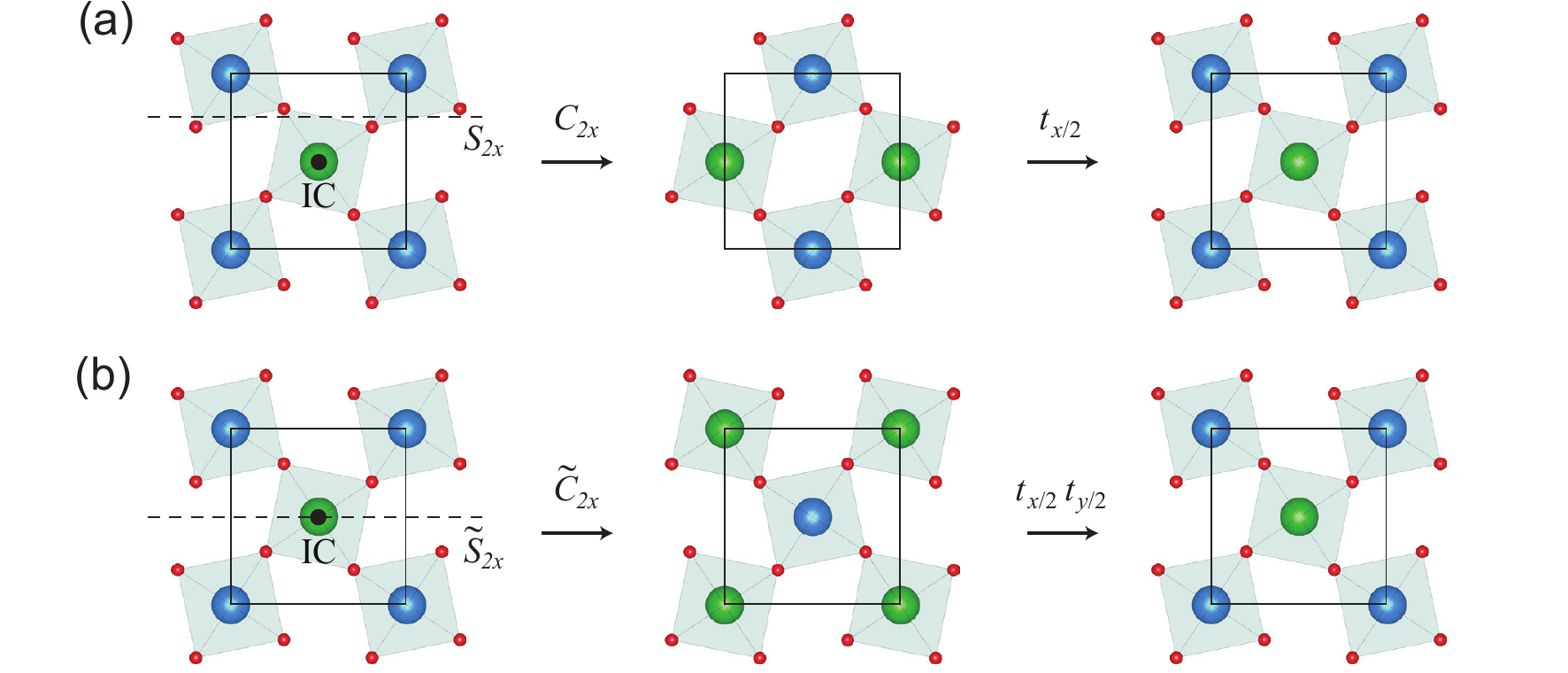} 
\caption{ 
\label{fig:Two NS}
(a) Off-centered screw rotation $S_{2x}=\{C_{2x}|\frac{1}{2}00\}=t_{x/2}C_{2x}$, where 
the rotation axis of $S_{2x}$ does not intersect the inversion center (IC).  
(b) Off-centered screw rotation $\widetilde{S}_{2x}=\{\widetilde{C}_{2x}|\frac{1}{2}\frac{1}{2}0\}=t_{x/2}t_{y/2}\widetilde{C}_{2x}$, where the rotation axis of $\widetilde{S}_{2x}$ intersects the inversion center.
} 
\end{figure}

In this section, we discuss the role of nonsymmorphic symmetry to the formation of essential Dirac semimetal state with and without d-wave order.
In the layer group, a nonsymmorphic symmetry $\{g|\vec{t}\}$ such as a glide mirror or a screw rotation is a combination of a point group operation $g$ and a half translation $\vec{t}$.
For our system (layer group p$4/mbm$), we have two orthogonal screw rotations $S_{2x}$ and $S_{2y}$, as shown in Fig. 1(b).
The screw rotation $S_{2x,2y}$ is composed of a rotation along the $x$-or $y$-axis and a half translation along that $x$-or $y$-axis.
We note that the rotation axis of $S_{2x,2y}$ does not intersect an inversion center, which plays a pivotal role to the protection of band degeneracies.
This kind of nonsymmorphic symmetry is called off-centered nonsymmorphic symmetry \cite{wieder_spin-orbit_2016_2,yang_topological_2017_2}.
One can see that the off-centered nonsymmorphic symmetry $S_{2x} =  t_{x/2}C_{2x}$ is equivalent to $\widetilde{S}_{2x} =  t_{x/2}t_{y/2}\widetilde{C}_{2x}$ as shown in Fig. \ref{fig:Two NS}.
Here,  $C_{2x}$ is a rotation operator of which axis is not located at the inversion center. 
Meanwhile, the axis of rotation operator $\widetilde{C}_{2x}$ is located at the inversion center, which makes the symmetry analysis mathematically simpler
because $\widetilde{C}_{2x}$ commutes with the inversion operator $P$.
Therefore, we will use $\widetilde{S}_{2x}$ instead of $S_{2x}$ in the subsequent mathematical evaluations.
We note that $(S_{2x})^2 = (\widetilde{S}_{2x})^2 = -e^{ikx}$ in the presence of SOC.
Similarly, $\widetilde{S}_{2y}$ $=$ $t_{x/2}t_{y/2}\widetilde{C}_{2y}$ can be used for the analysis of $S_{2y}$.

\subsubsection{Dirac line node in the absence of d-wave order}
In our system with strong spin-orbit coupling, the single off-centered screw rotation $S_{2x,2y}$ with inversion $P$ and time-reversal $\Theta$ guarantees the existence of Dirac point nodes (DPNs) at the BZ boundary.
An additional mirror symmetry $M_z$ extends the DPNs to a Dirac line node (DLN) along the $S_{2x,2y}$ invariant line.
Thus, two orthogonal off-centered screw rotations $S_{2x}$ and $S_{2y}$ with $P$, $\Theta$ and $M_z$ give rise to the symmetry protected DLN in the whole BZ boundary.
Let us discuss it step by step.

\begin{figure}[t]
\includegraphics[width=\textwidth]{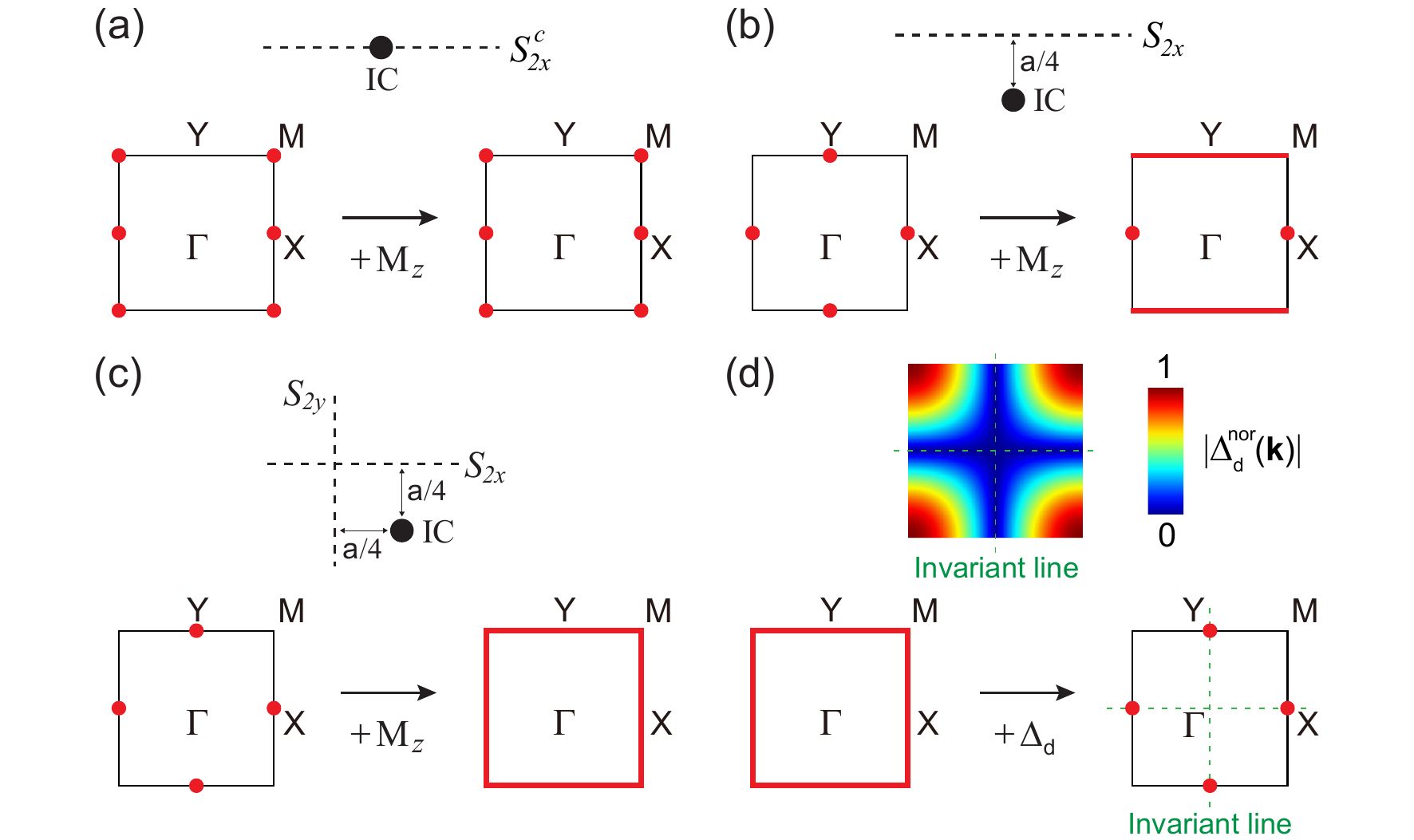} 
\caption{ 
\label{fig:Symmetry_summary}
(a) Conventional nonsymmorphic symmetry with mirror symmetry.
(b) Off-centered nonsymmorphic symmetry with mirror symmetry.
(c) Two orthogonal off-centered nonsymmorphic symmetries with mirror symmetry.
(d) Our system without and with d-wave order.
Red dots and lines denote DPN and DLN, respectively.} 
\end{figure}

First, let us see the role of the single off-centered screw rotation $S_{2x}$ in the presence of $P$ and $\Theta $ symmetries.
We now highlight the role of the off-centered nonsymmorphic symmetry $S_{2x}$ by comparing with a conventional nonsymmorphic symmetry $S_{2x}^c$.
The key difference between $S_{2x}$ and $S_{2x}^c$ is the offset of the rotation axis from the inversion center [see the top schematics in Figs. \ref{fig:Symmetry_summary}(a) and \ref{fig:Symmetry_summary}(b)], which gives different commutation relations with $P$.
Consequently, the positions of the Dirac point nodes are different for the two cases [see the left figures in Figs. \ref{fig:Symmetry_summary}(a) and \ref{fig:Symmetry_summary}(b)].
For the conventional nonsymmorphic symmetry $S_{2x}^c = t_{x/2}{C}_{2x}^c$, the rotation axis of ${C}_{2x}^c$ is located at the inversion center and hence $[{C}_{2x}^c,P]=0$.
Then the commutation relation with $P$ is given by
\begin{eqnarray}\label{seq:Con NS and P}
{S}_{2x}^cP=t_{x/2}{C}_{2x}^cP=Pt_{-x/2}{C}_{2x}^c=Pt_{-x}{S}_{2x}^c=e^{-ik_x}P{S}_{2x}^c.
\end{eqnarray}
In the last step, $t_{-x}$ operator acts on the eigenstate of ${S}_{2x}^c$.
Thus, only at $k_x=\pi$, $\{P,{S}_{2x}^c\}=0$.  
Combined with $P^2=1$, $\Theta^2=-1$, and $({S}_{2x}^c)^2=1$, 
the symmetry algebra gives rise to the fourfold-degenerate DPNs at the X and M points [see the left figure in Fig. \ref{fig:Symmetry_summary}(a)] \cite{wieder_spin-orbit_2016_2}.  That is, $\psi$, $P \psi$, $\Theta\psi$ and $P\Theta\psi$ form a set of fourfold-degenerate states where $\psi$ is an eigenstate of ${S}_{2x}^c$.
For the off-centered nonsymmorphic symmetry $S_{2x}$ or, equivalently, $\widetilde{S}_{2x} = t_{x/2} t_{y/2} \widetilde{C}_{2x}$,
the commutation relation is given by
\begin{eqnarray} \label{seq:comm_ns_inversion}
\widetilde{S}_{2x}P=t_{x/2}t_{y/2}\widetilde{C}_{2x}P=Pt_{-x/2}t_{-y/2}\widetilde{C}_{2x}=Pt_{-x}t_{-y}\widetilde{S}_{2x}=e^{-ik_x}e^{ik_y}P\widetilde{S}_{2x}.
\end{eqnarray}
Unlike equation (\ref{seq:Con NS and P}), there is another phase factor of $e^{ik_y}$,
which leads to $\{P, \widetilde{S}_{2x} \} = 0$ at the X and Y points.
Again, combined with $P\Theta$ symmetry, the symmetry algebra gives rise to DPNs at the X and Y points [see the left figure in Fig. \ref{fig:Symmetry_summary}(b)].
Note that $(\widetilde{S}_{2x})^2=1$ at the X point and $(\widetilde{S}_{2x})^2=-1$ at the Y point.

Next, let us discuss how an additional mirror symmetry $M_z$ extends the DPNs to a DLN along the $S_{2x}$ invariant line.
We are also going to use $\widetilde{S}_{2x}$ instead of $S_{2x}$ as mentioned before.
From the $P\Theta$ symmetry, all bands are doubly degenerate in the whole BZ.
For the formation of a DLN, two doubly degenerate bands should be degenerate and here $M_z$ plays such a role.
To see this,
we are going to analyse the $\widetilde{S}_{2x}$ eigenvalues of four states
\begin{eqnarray} \label{seq:four_states}
\ket{\pm,k_x,k_y},~~P \Theta \ket{\pm,k_x,k_y},~~M_z\ket{\pm,k_x,k_y},~~\text{and}~~M_zP\theta\ket{\pm,k_x,k_y}.
\end{eqnarray} 
Here, $\ket{\pm,k_x,k_y}$ is an eigenstate of $\widetilde{S}_{2x}$ that
satisfies the following eigenvalue equation
\begin{eqnarray}
\widetilde{S}_{2x}\ket{\pm,k_x,k_y} = \pm ie^{ik_{x}/2}\ket{\pm,k_x,k_y}
\end{eqnarray}
along the $\widetilde{C}_{2x}$ invariant lines ($k_y=0$ and $k_y=\pi$ lines).
In order to know the $\widetilde{S}_{2x}$ eigenvalues of three other states in equation (\ref{seq:four_states}), we evaluate the following two commutation relations.
The commutation relation between $\widetilde{S}_{2x}$ and $P\Theta$ is obtained as
\begin{eqnarray}\label{seq:comm_relation_1}
\widetilde{S}_{2x}P\Theta=e^{ik_x}e^{-ik_y}P \Theta\widetilde{S}_{2x}
\end{eqnarray}
by multiplying equation (\ref{seq:comm_ns_inversion}) by $\Theta$.
And the commutation relation between $\widetilde{S}_{2x}$ and $M_z$ is obtained by
\begin{eqnarray}\label{seq:comm_relation_2}
\widetilde{S}_{2x}M_z=t_{x/2}t_{y/2}\widetilde{C}_{2x}P\widetilde{C}_{2z}=-P\widetilde{C}_{2z}t_{x/2}t_{y/2}\widetilde{C}_{2x}=-M_z\widetilde{S}_{2x}.
\end{eqnarray}
Then, using equations (\ref{seq:comm_relation_1}) and (\ref{seq:comm_relation_2}),
one can obtain the following eigenvalue equations:
\begin{eqnarray}\label{seq:general_eigenvalu_ns}
\widetilde{S}_{2x} \left [ P\Theta\ket{\pm,k_x,k_y} \right ] &=& \mp i e^{ik_x/2} e^{ - ik_y} \left [ P\Theta\ket{\pm,k_x,k_y} \right ], \\
\widetilde{S}_{2x} \left [ M_z\ket{\pm,k_x,k_y} \right ] &=&   \mp ie^{ik_{x}/2}  \left [ M_z\ket{\pm,k_x,k_y} \right ], \\
\widetilde{S}_{2x} \left [ M_zP\theta\ket{\pm,k_x,k_y} \right ] &=& \pm ie^{ik_{x}/2}  e^{ - ik_y} \left [ M_z P\theta\ket{\pm,k_x,k_y} \right ].
\end{eqnarray}

 Now, let us discuss the $\widetilde{S}_{2x}$ eigenvalues of the four states  in equation (\ref{seq:four_states}).
Along the $k_y=\pi$ line, equation (\ref{seq:general_eigenvalu_ns}) implies
that a Kramers pair $\ket{\pm,k_x,k_y=\pi} $ and $ P\Theta \ket{\pm,k_x,k_y=\pi} $ share the same $\widetilde{S}_{2x}$ eigenvalue of $\pm ie^{ik_{x}/2}$. 
Also, equations (S8) and (S9) imply that a Kramers pair $M_z\ket{\pm,k_x,k_y=\pi} $ and $ M_zP\theta\ket{\pm,k_x,k_y=\pi} $ share the same $\widetilde{S}_{2x}$ eigenvalue of  $\mp ie^{ik_{x}/2}$.
We note that two Kramers pairs have the opposite $\widetilde{S}_{2x}$ eigenvalues, which permits the fourfold degeneracy of the four states listed in equation (\ref{seq:four_states}).
To sum up, the four states $\ket{\pm,k_x,k_y=\pi}$, $P\theta\ket{\pm,k_x,k_y=\pi}$, $M_z\ket{\pm,k_x,k_y=\pi}$ and $M_zP\theta\ket{\pm,k_x,k_y=\pi}$ with $\widetilde{S}_{2x}$ eigenvalues $\{\pm,\pm,\mp,\mp\}$ are degenerate and form a DLN along the $k_y=\pi$ line [see the right figure in Fig. \ref{fig:Symmetry_summary}(b)]. 

However, along the $k_y=0$ line, equation (\ref{seq:general_eigenvalu_ns}) means
that a Kramer pair $\ket{\pm,k_x,k_y=0} $ and $ P\Theta \ket{\pm,k_x,k_y=0}$ have the opposite $\widetilde{S}_{2x}$ eigenvalues.
Hence, when two doubly-degenerate bands meet, they anticross because a hybridization between the bands with the same $\widetilde{S}_{2x}$ eigenvalue is allowed. 
Therefore, DLN can not be stable along the $k_y=0$ line. 
Note that $M_z$ does not affect band degeneracy along the $k_y = 0$ line [see the right figure in Fig. \ref{fig:Symmetry_summary}(b)].
The reason is as follows.
The four states $\ket{\pm,k_x,k_y=\pi}$, $P\theta\ket{\pm,k_x,k_y=\pi}$, $M_z\ket{\pm,k_x,k_y=\pi}$, and $M_zP\theta\ket{\pm,k_x,k_y=\pi}$
have $\widetilde{S}_{2x}$  eigenvalues $\{ \pm,\mp,\mp,\pm \}$, respectively.
Thus, a hybridization between states with the same $\widetilde{S}_{2x}$ eigenvalue is allowed and hence $M_z$ does not give an additional degeneracy.

In the same way, the above argument used for the $k_y=0$ line can be applied to the conventional nonsymmorphic symmetry $S_{2x}^c$ because there is no $k_y$ dependence in the commutation relation as follows:
\begin{eqnarray}
S_{2x}^cP\Theta=e^{ik_x}P \Theta S_{2x}^c.
\end{eqnarray}
This commutation relation is the same with the commutation relation in equation (\ref{seq:comm_relation_1}) when $k_y=0$.
Thus, DLN can not be stable anywhere in the BZ for $S_{2x}^c$ [see the right figure in Fig. \ref{fig:Symmetry_summary}(a)]. 
Therefore, we can conclude that both the off-centered nature of $\widetilde{S}_{2x}$ and mirror $M_z$ protect the DLN along the BZ boundary parallel to the $\widetilde{S}_{2x}$ invariant line. 

Finally, we take into account an additional off-centered nonsymmorphic symmetry $S_{2y}$ [see the top schematic in Fig. \ref{fig:Symmetry_summary}(c)]. 
In the absence of $M_z$ symmetry, by carrying out the similar analysis into $S_{2y}$ like $S_{2x}$, 
one can find symmetry-protected DPNs at the X and Y points [see the left figure in Fig. \ref{fig:Symmetry_summary}(c)].   
We note that there is no change to the position of the symmetry-protected DPNs,
even if we have two orthogonal off-centered nonsymmorphic symmetries $S_{2x}$ and $S_{2y}$.
However, in the presence of $M_z$ symmetry,  
one can find a stable DLN along the $k_x=\pi$ line by performing the similar analysis into $S_{2y}$ like $S_{2x}$.
Therefore, we can conclude that the multiple symmetries as $S_{2x}$, $S_{2y}$, $M_z$ and $P\Theta$  protect
the DLN along the whole BZ boundary [see the right figure in Fig. \ref{fig:Symmetry_summary}(c)].

\subsubsection{Dirac point node in the presence of d-wave order}

In the presence of d-wave order,
among $S_{2x}$, $S_{2y}$, $M_z$ and $P\Theta$ symmetries, 
the nonsymmorphic symmetries $S_{2x}$ and $S_{2y}$ are broken due to the nature of hopping integral of d-SODW, 
which are explained in Section 1.3. 
However, we note that the d-SODW term
\begin{eqnarray}
H_\Delta= -4\Delta_{d} \sin(k_x/2)\sin(k_y/2)
\end{eqnarray}
vanishes along the $k_x=0$ and $k_y=0$ lines due to its d-wave form factor [see Fig. \ref{fig:Symmetry_summary}(d)]. 
For simplicity, we call these lines symmetry-invaraint lines where the nonsymmorphic symmetries are still maintained.
In the symmetry-invaraint lines, one can perform the similar symmetry analysis played by $S_{2x}$ and $S_{2y}$ as discussed in the section 1.2.1. 
Thus, the fourfold degeneracies along the whole BZ boundary are lifted, except for the points located at the symmetry-invariant lines,
which leads to a phase transition from DLN to DPN [Fig. \ref{fig:Symmetry_summary}(d)].
At the X and Y points, a set of $ \{ \ket{\pm,k_x,k_y}$, $P \ket{\pm,k_x,k_y}$, $\Theta\ket{\pm,k_x,k_y}$, $P\Theta\ket{\pm,k_x,k_y} \}$ forms a fourfold degeneracy.
In summary, we have applied the same argument in the section 1.2.1 to the protection of DPNs at the X and Y points located at the symmetry-invariant lines.
That is, the two Dirac point nodes at the X and Y points are protected by either $S_{2x}$ or $S_{2y}$ with  $P\Theta$ symmetry.

\subsection{Effect of d-wave order on the nonsymmorphic symmetry}

\begin{figure}[h]
\includegraphics[width=\textwidth]{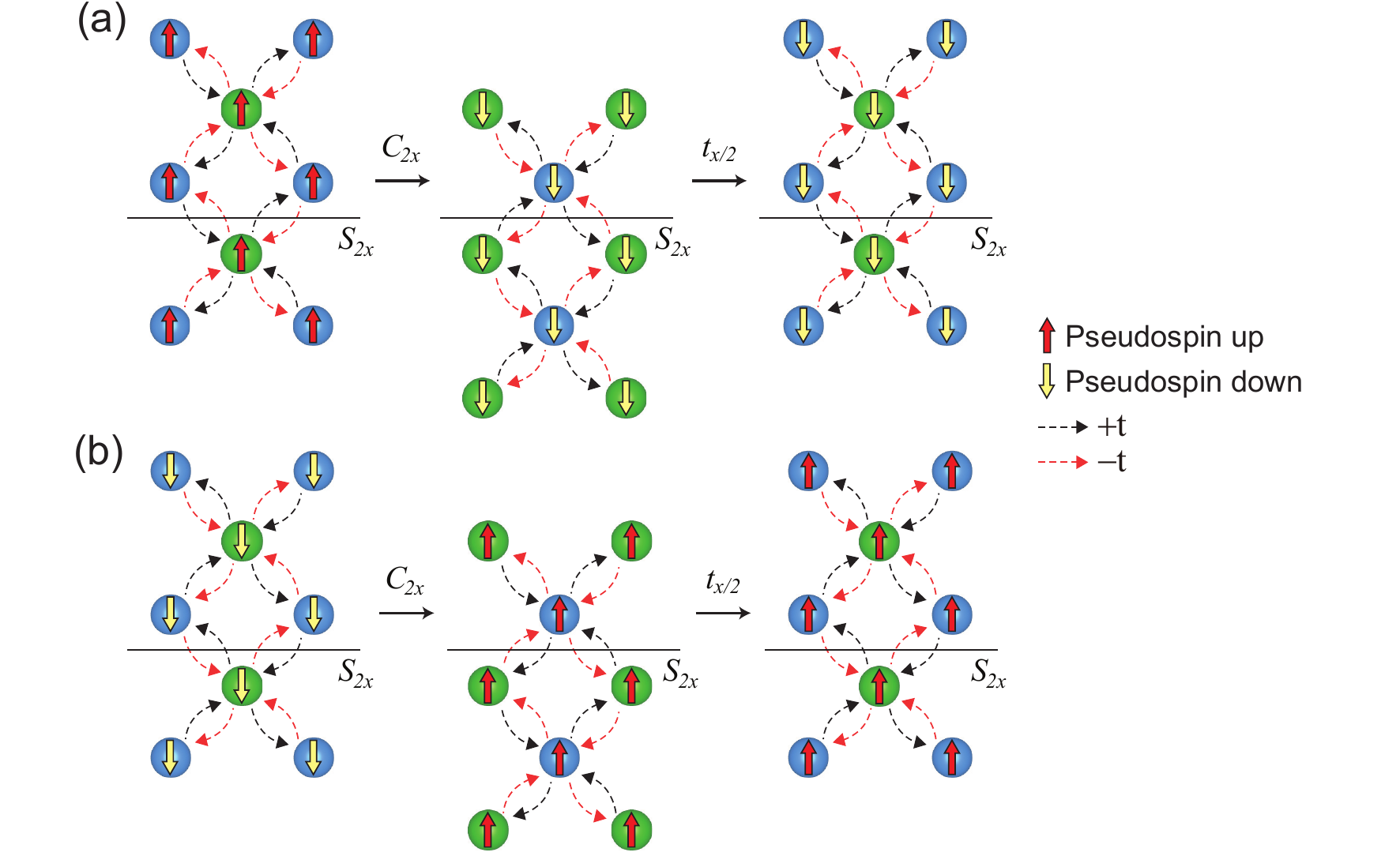} 
\caption{ 
\label{fig:NS_under_d}
(a),(b) Schematics for changes of configuration under $S_{2x}$ operation for (a) pseudospin up and (b) down states.
Black (red) dashed line denote the hopping integral of $+t$ ($-t$).} 
\end{figure}

In this section, we discuss the effect of d-wave electronic order on the crystalline symmetries. 
For d-SODW order, the hopping integrals for $J_{\text{eff}}=1/2$ pseudospin up and down states are shown in the left panel of Figs. \ref{fig:NS_under_d}(a) and \ref{fig:NS_under_d}(b), respectively. 
Note that the signs of hopping integrals between pseudospin up and down states are the opposite. 
Under $S_{2x}$ operation, as shown in Fig. \ref{fig:NS_under_d}, the configuration of hopping integrals for each pseudospin state does not convert into the original configuration of its opposite pseudospin state, which means the nonsymmorphic operation $S_{2x}$ is a no longer symmetry operation in the presence of d-SODW order. 
Similarly, one can consider $S_{2y}$ operation.
For other symmetries such as Mirror $M_z$, inversion $P$ and time-reversal $\Theta$, one can find that they are not broken in the presence of d-SODW order.
Therefore, we can conclude that only nonsymmorphic symmetries $S_{2x}$ and $S_{2y}$ are broken due to the nature of hopping integrals of d-SODW order. 

\section{\normalsize M\lowercase{odel }H\lowercase{amiltonian}}

\subsection{Tight-binding Hamiltonian}

With all five d orbitals, we constructed the 2D TB+SOC Hamiltonian $H_0$, which is given by
\begin{eqnarray}
H_0= \sum_{ij,\mu\nu,\sigma} t_{ij}^{\mu\nu,\sigma} c_{i\mu\sigma}^{\dagger} c_{j\nu\sigma} 
+ \sum_{i,\mu,\sigma} \epsilon_{\mu} c_{i\mu\sigma}^{\dagger} c_{i\mu\sigma} 
+ \sum_{i,\mu\nu,\sigma\sigma'} \lambda_{SOC} \bra{\mu}\mathbf{L}\ket{\nu} \cdot \bra{\sigma}\mathbf{S}\ket{\sigma'} c_{i\mu\sigma}^{\dagger} c_{i\nu\sigma'},
\end{eqnarray}
where $c_{i\mu\sigma}^{\dagger}$ is a creation operator for an electron with spin $\sigma$ in the $\mu$th orbital at site $i$,
$\lambda_{SOC}$ is a SOC parameter, 
and $t_{ij}^{\mu\nu,\sigma}$ is the complex hopping integrals between sites $i$ and $j$ of up to the fifth nearest neighbors 
as given in the Ref. \cite{zhou_correlation_2017_2}.
The crystalline electric field effects were taken into account in the on-site energy term $\epsilon_{\mu =( d_{yz}, d_{zx}, d_{xy}, d_{3z^2-r^2}, d_{x^2-y^2})} = (0,0,202,3054,3831)$ meV.
Here, $\mathbf{L}$ and $\mathbf{S}$ are the orbital and spin angular momentum operators, respectively.

For the electron correlation effect, we considered the five-orbital Hubbard Hamiltonian $H_U$ given by, 
\begin{align}
\begin{split}
H_U  = U\sum_{i,\mu} \hat{n}_{i\mu\uparrow} \hat{n}_{i\mu\downarrow}
+ (U'-J/2)\sum_{i,\mu<\nu} \hat{n}_{i\mu} \hat{n}_{i\nu}
 \\
-J \sum_{i,\mu\neq\nu} \mathbf{S}_{i\mu} \cdot \mathbf{S}_{i\nu}
+ J \sum_{i,\mu\neq\nu} c_{i\mu\uparrow}^{\dagger} c_{i\mu\downarrow}^{\dagger} c_{i\nu\downarrow} c_{i\nu\uparrow},
\end{split}
\end{align}
where $\hat{n}$ is a density operator, $U$ and $U'$ are the local intraorbital and interorbital Coulomb repulsions, respectively, and $J$ is the Hund's rule coupling with $U=U'+2J$. 
For our Tb-doped system, we used $U=1.6$ eV on the Ir site which is a calculated value in the undoped system using constrained random-phase approximation (cRPA) \cite{liu_electron_2016}.
As pointed out for parent Sr$_2$IrO$_4$ compound \cite{zhou_correlation_2017_2}, 
we note that the renormalization of SOC due to the $U$ is also significant in our Tb-doped system as $\lambda_{SOC}^{\text{eff}} \sim 786$ meV. 
We used $J=0$ because our doped systems are paramagnetic. 

To account for the observed pseudogap, we introduced d-SODW order as a symmetry-breaking order compatible with the symmetry of our doped systems.
It was suggested to have an electronic origin arising from the intersite Coulomb interaction \cite{zhou_correlation_2017_2}.
The d-SODW Hamiltonian $H_\Delta$ in the $J_{\text{eff}}=1/2$ basis is given by 
\begin{eqnarray}
H_\Delta= i\Delta_d \sum_{i\in \text{Ir}_\text{A},\sigma = \pm} \sum_{j=i+\delta} (-1)^{i_y+j_y} \sigma \gamma_{i,\sigma}^{\dagger} \gamma_{j,\sigma}
+ h.c.,
\end{eqnarray}
where $\Delta_d$ is a d-SODW parameter, $\delta=\pm\hat{x},\pm\hat{y}$, $(-1)^{i_y+j_y}$ is the nearest-neighbor d-wave form factor, 
and $\gamma_{\pm}^{\dagger}\ket{0}=\ket{J_{\text{eff}}=1/2,J_z=\pm1/2}$ from which $\gamma_{\sigma} = 1/\sqrt{3}(i\sigma d_{yz,\bar{\sigma}}-d_{zx,\bar{\sigma}}+id_{xy,\sigma})$.
To reproduce ARPES spectra, we used $\Delta_d=80$ meV and introduced small amount of electrons ($0.01e/\text{Ir}$ atom) at Ir site. 
Although it does not seem to be compatible with the experimental result that Tb doping is isovalent \cite{wang_decoupling_2015_2}, 
inherent oxygen deficiency can alter the valency of Ir atom \cite{sung_crystal_2016}, allowing some variations of the valency of Ir atom. 
The calculated optical properties based on the calculated band structure with above parameters also agree well with the THz result (see section 7.5).

Therefore, the total Hamiltonian $H$ is 
\begin{align}
\begin{split}
H & = H_0+H_U+H_\Delta 
 \\
& =  \sum_{ij,\mu\nu,\sigma} t_{ij}^{\mu\nu,\sigma} c_{i\mu\sigma}^{\dagger} c_{j\nu\sigma} 
+ \sum_{i,\mu,\sigma} \epsilon_{\mu} c_{i\mu\sigma}^{\dagger} c_{i\mu\sigma} 
+ \sum_{i,\mu\nu,\sigma\sigma'} \lambda_{SOC} \bra{\mu}\mathbf{L}\ket{\nu} \cdot \bra{\sigma}\mathbf{S}\ket{\sigma'} c_{i\mu\sigma}^{\dagger} c_{i\nu\sigma'} 
\\
&\quad + U\sum_{i,\mu} \hat{n}_{i\mu\uparrow} \hat{n}_{i\mu\downarrow}
+ U\sum_{i,\mu<\nu} \hat{n}_{i\mu} \hat{n}_{i\nu}
+ [i\Delta_d \sum_{i\in \text{Ir}_\text{A},\sigma=\pm} \sum_{j=i+\delta} (-1)^{i_y+j_y} \sigma \gamma_{i,\sigma}^{\dagger} \gamma_{j,\sigma}
+ h.c.].
\end{split}
\end{align}

\subsection{Effective Hamiltonian analysis}

In this section, we construct the low-energy effective Hamiltonian in the $J_{\text{eff}}=1/2$ basis.
We provide the form of symmetry operators in the $J_{\text{eff}}=1/2$ basis and discuss the symmetry of the effective Hamiltonian.
Then, using k$\cdot$p theory and unitary transformation, we construct Dirac Hamiltonians for DPN state near the X and Y points, which will be used in optical conductivity calculations in Section 7.

\subsubsection{Effective Hamiltonian and symmetry operators in the $J_{\text{eff}}=1/2$ basis}
The effective Hamiltonian of Sr$_2$IrO$_4$ in the $J_{\text{eff}}=1/2$ basis is given by \cite{carter_theory_2013,zhou_correlation_2017_2} 
\begin{eqnarray} \label{seq:effective H}
H_{\text{eff}} (\mathbf{k}) = \epsilon_1 (\mathbf{k}) 
+ \epsilon_2 (\mathbf{k}) \tau_x 
+  \epsilon_3 (\mathbf{k}) \tau_y \sigma_z 
+ \epsilon_\Delta (\mathbf{k}) \tau_y \sigma_z,
\end{eqnarray}
where 
\begin{align}
\begin{split}
\epsilon_1 (\mathbf{k}) & =  
2t_1\left [ \cos ( k_x ) + \cos( k_y ) \right ] + 4  t_{1p}  \cos( k_x   )  \cos ( k_y  ) ,
\\
\epsilon_2 (\mathbf{k}) & =  4 t_2 \cos( k_x /2) \cos( k_y /2),
\\
\epsilon_3 (\mathbf{k}) & =  4 t_3  \cos( k_x/2) \cos( k_y/2),
\\
\epsilon_{\Delta} (\mathbf{k}) & =   -4\Delta_d \sin( k_x/2) \sin( k_y/2).
\end{split}
\end{align}
Here, $t_1, t_{1p}, t_2 $, and $t_3 $ are hopping integrals, 
$\Delta_d$ is the d-wave order parameter, and
$\tau_i$'s and $\sigma_i$'s $(i = x, y, z)$ are Pauli matrices in the sublattice and $J_{\text{eff}}=1/2$ basis, respectively.
When  $\Delta_d=0$ ($\Delta_d \neq 0 $), this effective Hamiltonian $H_{\text{eff}}$ describes DLN (DPN) state.
Each parameter can be obtained by fitting with the calculated band dispersion using the five-orbital tight-binding model. 

Essential symmetry operators can be represented in the $J_{\text{eff}}=1/2$ basis as
%
\begin{align} \label{seq:all symmetry operators}
\begin{split}
 \Theta & =  i \sigma_y K \otimes (\mathbf{k} \rightarrow -\mathbf{k}),
 \\
 P & =  I \otimes (\mathbf{k} \rightarrow -\mathbf{k}),
 \\
 M_z & =  i \sigma_z \otimes (k_z \rightarrow -k_z), 
 \\
 S_{2x} & =  i \tau_x \sigma_x \otimes (k_y \rightarrow -k_y),
 \\
 S_{2y} & =  i \tau_x \sigma_y \otimes (k_x \rightarrow -k_x),
\end{split}
\end{align}
where $K$ is the complex conjugation operator. In the absence of d-wave order $(\Delta_d = 0)$, effective Hamiltonian is invariant under all symmetry operations in Eq. (\ref{seq:all symmetry operators}):
\begin{align}\label{seq:symmetry_relations}
\begin{split}
 P H_{\text{eff}}(\mathbf k) P^{-1} & =  H_{\text{eff}}(-\mathbf k ),
 \\
 \Theta H_{\text{eff}}(\mathbf k) \Theta^{-1} & =  H_{\text{eff}}(-\mathbf k ),
 \\
 M_z H_{\text{eff}}(k_x,k_y,k_z) M_z^{-1} & =  H_{\text{eff}} (k_x,k_y,-k_z),
 \\
 S_{2x} H_{\text{eff}}(k_x,k_y,k_z) S_{2x}^{-1} & =  H_{\text{eff}} (k_x,-k_y,k_z),
 \\
 S_{2y} H_{\text{eff}}(k_x,k_y,k_z) S_{2y}^{-1} & =  H_{\text{eff}} (-k_x,k_y,k_z),
\end{split}
\end{align}
where parity operations such as $(k_x \rightarrow -k_x)$ of symmetry operators in Eq. (\ref{seq:all symmetry operators}) only acted on the right side in Eq. (\ref{seq:symmetry_relations}).
However, in the presence of d-wave order $(\Delta_d \neq 0)$,
the symmetry relations for the two screw rotations $S_{2x}$ and $S_{2y}$ in Eq. (\ref{seq:symmetry_relations}) do not hold, except for the invariant lines $k_x =0$ and $k_y=0$, which is consistent with the previous symmetry analysis in Section 1.2.2.

\subsubsection{Dirac Hamiltonian for DPN state}
When d-wave order is nonzero, there exist the Dirac point nodes at the X and Y points as shown in the Fig. 1(e).
To obtain explicit form of Dirac Hamiltonian, using k$\cdot$p theory, we expand the effective Hamiltonian $H_{\text{eff}}(\mathbf k)$ around the $\text{X} = (\pi, 0)$.
By the change of variables $k_x \rightarrow  \pi + k_x$ and $k_y \rightarrow   k_y$,
we get
\begin{align}
H_{\text{eff}}  =   -2 t_2 k_x \tau_{x} -2 ( t_3 k_x +\Delta_d  k_y  ) \tau_{y} \sigma_{z}.
\end{align}
Next, we use the following the coordinate transformation:
\begin{align}
k_x'= k_x \cos \theta  + k_y \sin \theta , ~~~~~~k_y' = k_y \cos \theta  -  k_x \sin \theta,
\end{align}
where $\theta =  \frac{1}{2} \tan^{-1}\left(\frac{2 t_3 \Delta_d }{t_2^2 + t_3 ^2 - \Delta_d^2} \right)$.
Then we apply the unitary transformation $H' = U^{\dagger} H U$ with a unitary operator is $U = \exp [ - i \phi \tau_z /2]$, where
$\phi  = \tan^{-1} \left ( \frac{ t_3 \cos \theta + \Delta_d \sin \theta}{t_2 \cos \theta } \right )$.
Then we obtain the following anisotropic Dirac Hamiltonian:
\begin{eqnarray} 
H' = -  \alpha k_x' \tau_x - \beta k_y' \tau_y \sigma_z,
\end{eqnarray}
where 
$\alpha^2  =  2  [ \Delta_d^2 + t_2^2 + t_3^2 + (t_2^2 + t_3^2-\Delta_d^2) \cos 2 \theta +   2 t_3 \Delta_d \sin 2 \theta   ]$ and 
$\beta^2  =  2 [ \Delta_d^2 + t_2^2 + t_3^2 - (t_2^2 + t_3^2-\Delta_d^2) \cos 2 \theta -   2 t_3 \Delta_d \sin 2 \theta ]$.
Then energy eigenvalues are given by
\begin{eqnarray}
\epsilon(\mathbf k) & = &  
\pm  \sqrt{
         \alpha^2 k_x'^2 + \beta^2 k_y'^2
        }.
\end{eqnarray}
Similarly, the anisotropic Hamiltonian near the another Dirac point node at $\text{Y} = (0, \pi)$ is given by
\begin{eqnarray} \label{seq:DPN effective H}
H' = -  \beta k_x' \tau_x - \alpha k_y' \tau_y \sigma_z,
\end{eqnarray}
where eigenvalues are given by
\begin{eqnarray}
\epsilon(\mathbf k) & = &  
\pm  \sqrt{
         \beta^2 k_x'^2 + \alpha^2 k_y'^2
        }
,
\end{eqnarray}
where an anisotropic factor is given by $ \eta=\frac{\alpha}{\beta} \approx 2.45$.
Note that, in this low-energy limit, two Dirac cones are approximately related by the $90^{\circ}$ rotation.

Because of the mirror symmetry $M_z  =  i \sigma_z \otimes (k_z \rightarrow -k_z)$,
the effective Hamiltonian at the Y point for DPN state in Eq. (\ref{seq:DPN effective H}) can be divided into two sub-Hamiltonians according to the mirror eigenvalues
$\lambda = \pm i $:
\begin{eqnarray}
H_{\text{DPN}}^{\pm} = - v_F \hbar \left ( k_x \tau_x \pm \eta  k_y \tau_y  \right ),
\end{eqnarray}
where $v_F=\beta / \hbar $.
To make the problem easier, we use the following polar coordinates $(k, \theta_{k})$:
\begin{eqnarray}\label{seq:polar_coordinates}
k_x &  = &  k  \cos \theta_k,
\\
k_y &  = &  \frac{k}{\eta} \sin \theta_k.
\end{eqnarray}
Then the Hamiltonian is further transformed into the simpler form, i.e., the form of Dirac Hamiltonian:
\begin{eqnarray}
H_{\text{DPN}}^{\pm}  = 
- v_F \hbar k
\left(\begin{array}{c c}
0 & e^{\mp i \theta_k} \\
e^{ \pm i \theta_k}  & 0
\end{array}\right)
.
\end{eqnarray}
Here, the energy eigenvalue is given by $E_{s \mathbf k} = s v_F \hbar \sqrt {k_x^2 + \eta^2 k_y^2} = s v_F \hbar k$,
where $s=+1$ and $-1$ denote the conduction and valence bands, respectively.
The corresponding eigenstate is given by
\begin{eqnarray} \label{seq:wave_fun}
u_{s \mathbf k}^{\pm} = \frac{1}{ \sqrt2 }
\left(\begin{array}{c}
e^{\mp i \theta_k } \\ - s 
\end{array}\right) .
\end{eqnarray}
The velocity
$ v_{\mathbf k}^{(i)} = \frac{1}{\hbar} \frac{ \partial E_{s \mathbf k}}{\partial k_{i}} $
can be expressed as
\begin{eqnarray}
v^{(x)}_{\mathbf k} &  = &  s  v_F \cos \theta_k,
\\
v^{(y)}_{\mathbf k} &  = &  s \eta v_F \sin \theta_k.
\end{eqnarray}

\section{\normalsize S\lowercase{ample preparation}}

The single crystals studied were grown using flux growth methods from off-stoichiometric quantities of SrCO$_3$, IrO$_2$ and SrCl$_2$.
For Sr$_2$(Ir$_{0.97}$Tb$_{0.03}$)O$_4$ \cite{wang_decoupling_2015_2} and (Sr$_{0.945}$La$_{0.055}$)$_2$IrO$_4$ \cite{chen_influence_2015_2}, Tb$_4$O$_7$ and La$_2$O$_3$ were appropriately included as starting materials, respectively.

\section{\normalsize ARPES \lowercase{results}}

\subsection{Experimental details}

ARPES measurements were performed at the beamline 4.0.3 (MERLIN) of the Advanced Light Source, Lawrence Berkeley National Laboratory. 
Spectra were acquired with VG-Scienta R8000 electron analyser. 
The single crystals were cleaved in-situ and data were taken at 50 and 100 K with a vacuum better than $5 \times 10^{-11}$ Torr. 
Total energy resolution was set to 15 meV at photon energies of $h\nu = 70 ~\text{and}~ 80$ eV.

\subsection{Linearity of Dirac band}

\begin{figure}[h]
\includegraphics[width=\textwidth]{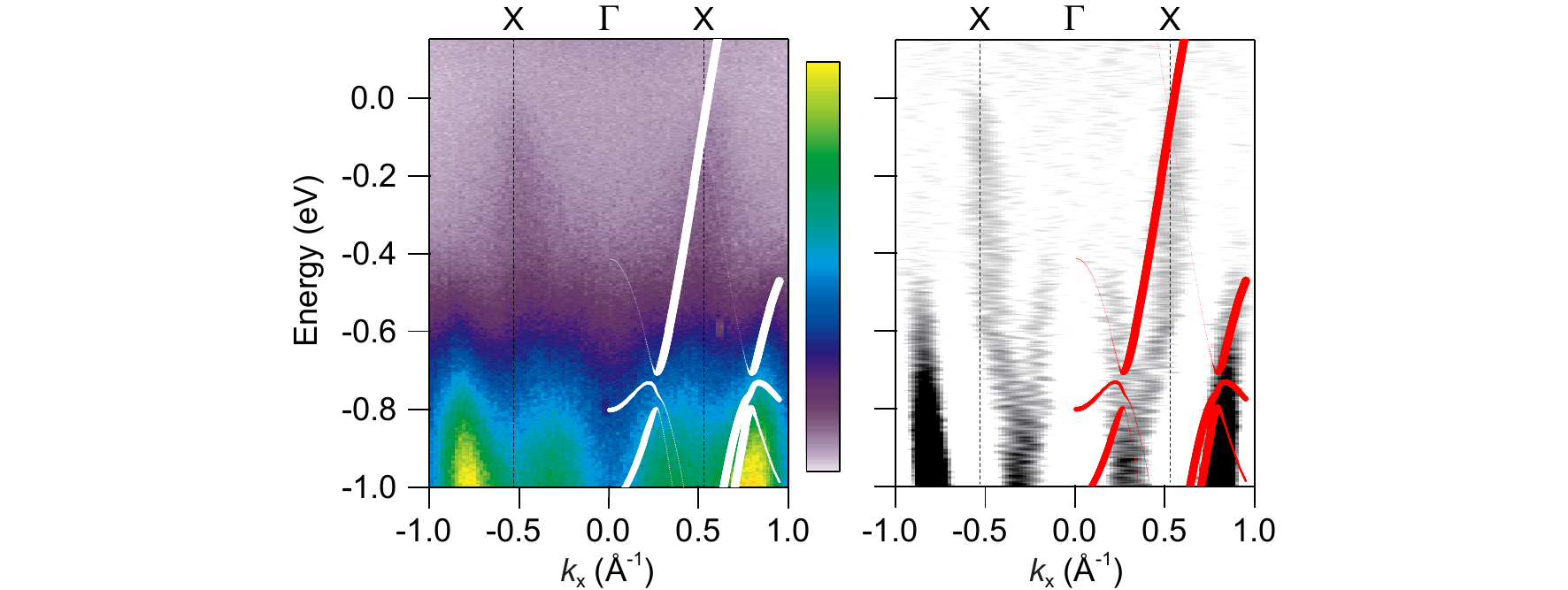}
\caption{
\label{fig:ARPES_unfolding}
Measured ARPES spectra for Sr$_2$(Ir$_{0.97}$Tb$_{0.03}$)O$_4$ at photon energy $h\nu=70$ eV and temperature $T=50$ K.  Second derivative of ARPES intensity plot is shown. The ARPES data are overlaid with calculated unfolded band structures (white and red lines) where the thickness of lines represents unfolding weights.
}
\end{figure}

Linear band dispersion is one of the necessary conditions to define Dirac band, and thus we discuss linearity of Dirac band in this section based on additional ARPES results. 
Figure \ref{fig:ARPES_unfolding} displays ARPES spectra of Sr$_2$(Ir$_{0.97}$Tb$_{0.03}$)O$_4$ along the X$-\Gamma-$X direction which was taken at photon energy 70 eV.
Measured electronic structure exhibits unfolded linear dispersion, and it is more clearly seen in the second derivative data.
It is consistent with the linear band obtained at photon energy 80 eV in Fig. 2(e) in the main text.
Also, both ARPES data are in good agreement with the calculated band structure including linear band dispersion as well as parabolic band at the $\Gamma$ point. 
By the way, the invisibility of the folded bands have been also observed in previous electronic structure studies on electron-doped Sr$_2$IrO$_4$ \cite{kim_fermi_2014_2,de_la_torre_collapse_2015_2,kim_observation_2016_2}.
Especially, for La-doped samples, it is reported that the folding of spectral weight of linear band is highly photon energy dependent \cite{de_la_torre_collapse_2015_2}.
The reason why counterpart of the linear band is not seen is attributed to the disorder effect of dopants and strong electron correlation, which are main sources to make the ARPES spectrum broaden out.
Accounting for the unfolding effect, we perform the band-unfolding calculation \cite{boykin_practical_2005}.
Indeed, we find that the calculated unfolded band structure is consistent with ARPES data as shown in Fig. \ref{fig:ARPES_unfolding}.
Comparing ARPES data and band structure calculations, we estimate the binding energy of Dirac point as $E_{\text B} \sim 50$ meV
and conclude that the linear Dirac band can be identified in Sr$_2$(Ir$_{0.97}$Tb$_{0.03}$)O$_4$.




\subsection{Pseudogap feature}

\begin{figure}[h]
\includegraphics[width=\textwidth]{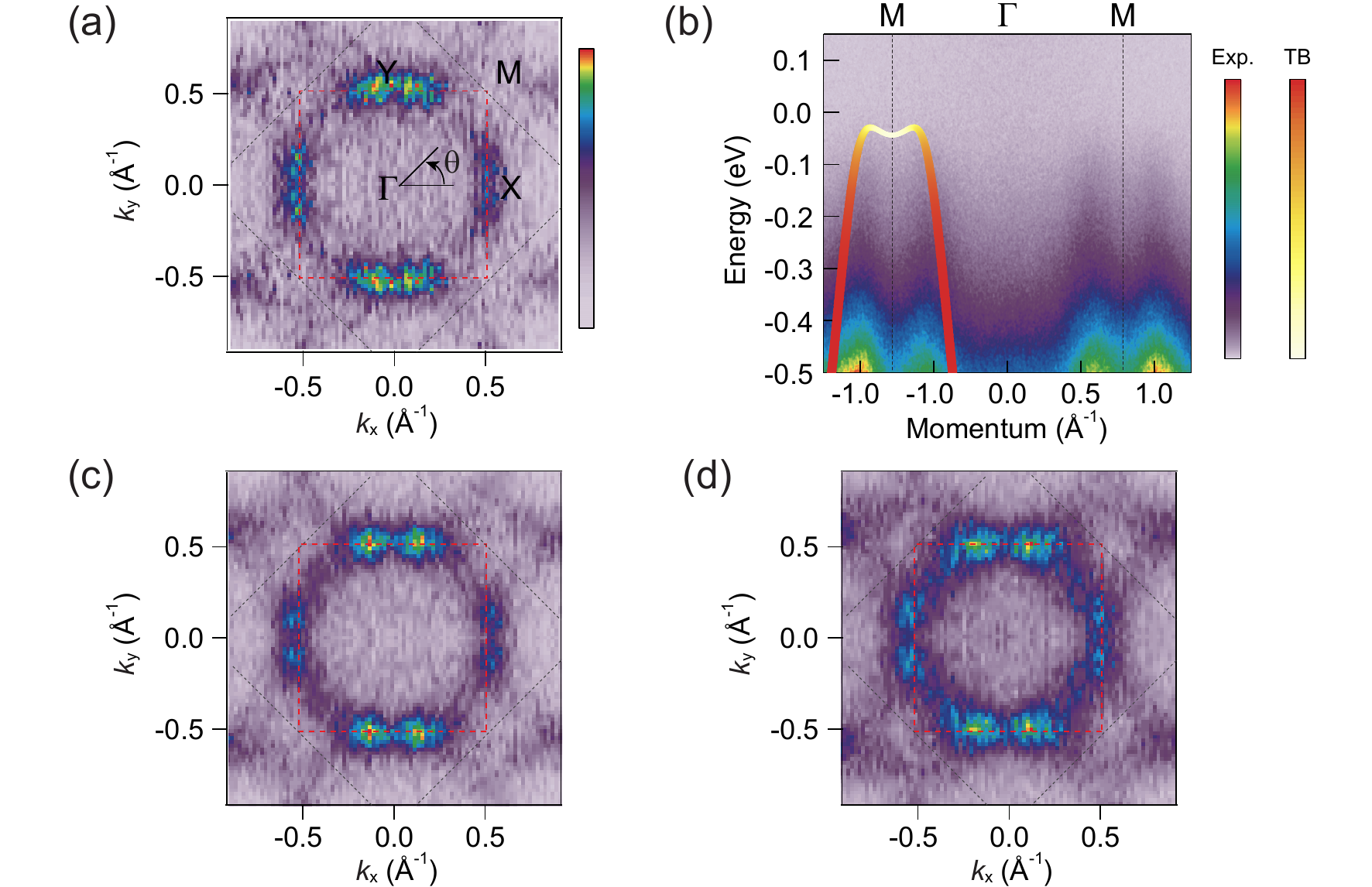}
\caption{
\label{fig:ARPES_unfolding2}
(a) Constant binding energy $k_{\text x}-k_{\text y}$ map at $E_{\text B}=E_{\text F}$, 
(b) ARPES spectra along the $\text{M}-\Gamma-\text{M}$ line, and
(c),(d) Constant binding energy $k_{\text x}-k_{\text y}$ maps at (c) $E_{\text B}=0.1$ and (d) $E_{\text B}=0.2$ eV of Sr$_2$(Ir$_{0.97}$Tb$_{0.03}$)O$_4$ at photon energy $h\nu=80$ eV and temperature $T=100$ K.
In (b), ARPES spectra are overlaid with calculated unfolded tight-binding (TB) band structures where the color bar indicates the unfolding weight.
}
\end{figure}

The d-wave pseudogap has been observed in electron-doped iridates \cite{kim_fermi_2014_2,de_la_torre_collapse_2015_2,kim_observation_2016_2}.
Similar to electron doping, we find that the isovalent Tb doping also gives rise to the d-wave pseudogap [Fig. \ref{fig:ARPES_unfolding2}(a)].
Note that DLN should exhibit hole pocket at the M point [Fig. 2(a)], which is not consistent with the present data.
In Fig. \ref{fig:ARPES_unfolding2}(b), the gapped band structure around the M point is clearly shown, which is consistent with the band structure of DPN state in Fig. 2(b).
The gapped band structure is more easily seen in our constant binding energy $k_{\text x}-k_{\text y}$ maps [Figs. \ref{fig:ARPES_unfolding2}(a), \ref{fig:ARPES_unfolding2}(c), and \ref{fig:ARPES_unfolding2}(d)]. 
As the binding energy $E_{\text B}$ increases, the angle-dependent ($\theta$) intensity variation becomes reduced, which is consistent with the case of d-wave pseudogap.
By the way, unlike tight-binding band structure in Fig. 2(b), the valence band top is not clearly seen in ARPES spectra [Fig. \ref{fig:ARPES_unfolding2}(b)], which is similar with La-doped sample \cite{de_la_torre_collapse_2015_2}.
As one of the possible explanations, we calculate band structure near the M point by considering unfolding effect.
It shows weak intensity near the valence band top, which is consistent with the ARPES data [Fig. \ref{fig:ARPES_unfolding2}(b)].

\section{\normalsize D\lowercase{etails of} TH\lowercase{z experiments}}

\subsection{Introduction of emitter-sample hybrid terahertz-time-domain spectroscopy (ES-TDS)}

Unbiased low-temperature-grown (LT) GaAs was illuminated by femtosecond laser pulses with 800 nm wavelength and 80 MHz repetition rate. 
Photo-excited carriers were accelerated by a built-in field near the surface 
and a THz wave was generated via a surge current mechanism \cite{gu_study_2002}. 
By using a pair of off-axis parabolic mirrors, 
a THz wave was delivered and focused to the photoconductive antenna by which the temporal profile of the THz electric field was acquired. 
Importantly, the sample was located at the back side of the THz emitter and the THz wave travelling into the emitter was reflected back by the emitter-sample interface \cite{han_extraction_2014}. 
The reflected THz wave was recorded in the same time domain as the THz wave emitted directly into the air, 
and it appeared with a 11 ps time delay corresponding to an additional travel time of the THz wave in the emitter medium. 
Using the Fourier transform technique, the magnitude and phase spectra of each THz wave were obtained. 
Complex optical constants, including optical conductivity spectra, were extracted by solving the Fresnel equation. 
For the low $T$ measurements, the sample was mounted in a liquid-nitrogen-cooled cryostat. 
Optical constants were determined at each $T$ with a proper consideration of the $T$-dependent optical constants of the emitter medium. 

\subsection{Advantanges of ES-TDS for measuring small-sized samples}

\begin{figure}[b]
\includegraphics[width=\textwidth]{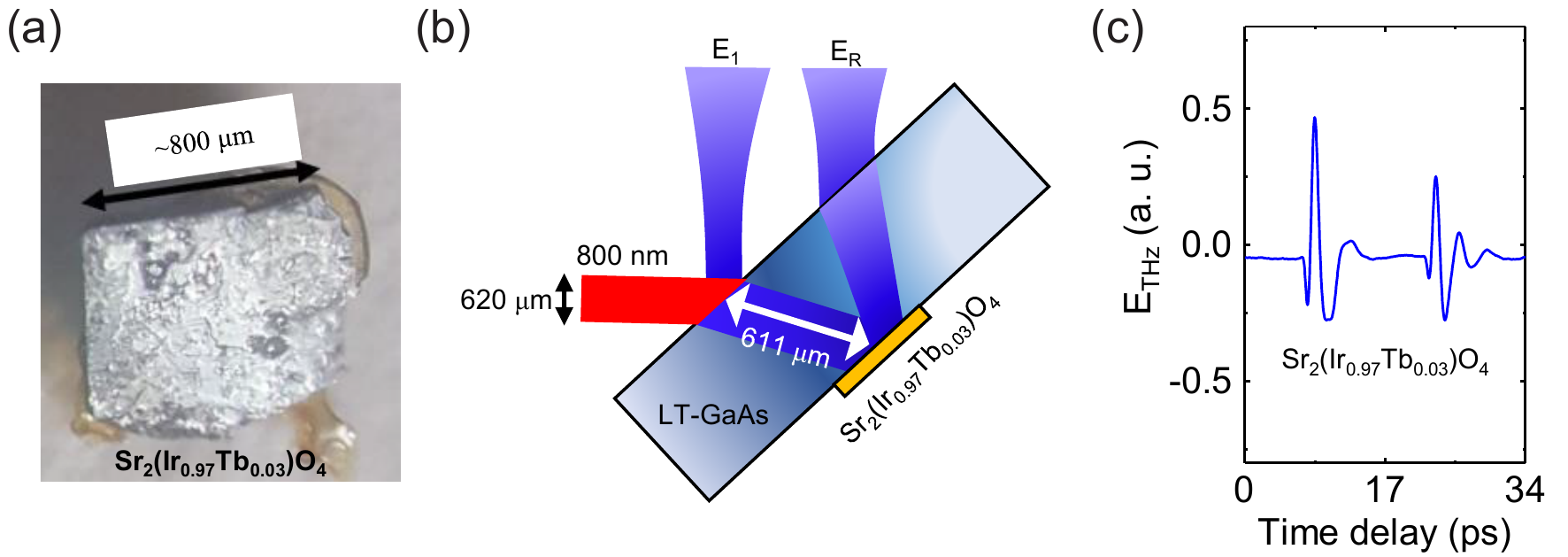} 
\caption{ 
\label{fig:THz1}
Optical image of the Sr$_2$(Ir$_{0.97}$Tb$_{0.03})$O$_4$ crystal and concept of emitter-sample hybrid THz time-domain spectroscopy.
(a) Picture of the Sr$_2$(Ir$_{0.97}$Tb$_{0.03}$)O$_4$ crystal used for the THz experiments. 
(b) Schematic of the beam paths. 
Width of pumping laser (800 nm wavelength) is about 620 $\mu$m and a traveling length of THz wave inside the LT-GaAs (THz emitter) is 611 $\mu$m. 
E$_1$ denotes the generated THz pulse propagating into the air. 
E$_\text{R}$ indicates the reflected THz pulse by the Sr$_2$(Ir$_{0.97}$Tb$_{0.03}$)O$_4$.
(c) THz pulses of E$_1$ and E$_\text{R}$ recorded in the time-domain.} 
\end{figure}

The Sr$_2$(Ir$_{1-x}$Tb$_x$)O$_4$ and (Sr$_{1-y}$La$_y$)$_2$IrO$_4$ single crystals we investigated are tiny, for example, only 0.8 mm wide for $x=0.03$ [see Fig. \ref{fig:THz1}(a)] and 0.7 mm wide for $y=0.055$, 
and hence it is technically difficult to obtain a reliable result about the optical conductivity spectra in the THz frequency range where the wavelength is comparable to the sample size. 
In this work, we employed an improved THz spectroscopy technique, 
i.e., ES-TDS \cite{han_extraction_2014,han_application_2017,han_radiating_2018}. 
The main strategy is to locate the sample at the back side of the THz emitter, LT-GaAs, where the THz wave is generated at the surface via a surge-current mechanism [see Fig. \ref{fig:THz1}(b)] \cite{gu_study_2002}. 
This provides us with two important advantages. 
First, we easily fit the THz beam into the sample area by adjusting the size of the pumping laser beam \cite{han_radiating_2018}. For example, with a 620 $\mu$m wide pumping beam, the THz wave beam size can be only 670 $\mu$m at the sample position; the sample is located within the Rayleigh length \cite{han_characteristics_2019}, 
and a slight larger THz beam size than the pumping beam size is simply due to a grazing incidence geometry. 
Second, we can make a reference correction for the Fresnel equation more conveniently by comparing the reflected beam from the sample (E$_\text{R}$) with the reference beam emitted directly into the air [E$_1$ in Fig. \ref{fig:THz1}(b)]. 
Since they are obtained by a single scan in the time-domain where each response appears with an enough time-interval as shown in Fig. \ref{fig:THz1}(c) \cite{han_extraction_2014}. 
This allows us to obtain more reliable results when the measurement is done as a function of $T$ \cite{han_application_2017}. 
Meanwhile, as can be seen in Fig. \ref{fig:THz1}(a), Sr$_2$(Ir$_{0.97}$Tb$_{0.03})$O$_4$ has step-like and/or crater-like defects. Their sizes are mostly about 10 $\mu$m, and surely smaller than 100 $\mu$m even for the largest defects. Therefore, we can assume that THz light should be specularly reflected from the sample surface with a minimal influence from the rough surface.

\subsection{Characterization of the THz beam profile at the sample position}

\begin{figure}[h]
\includegraphics[width=\textwidth]{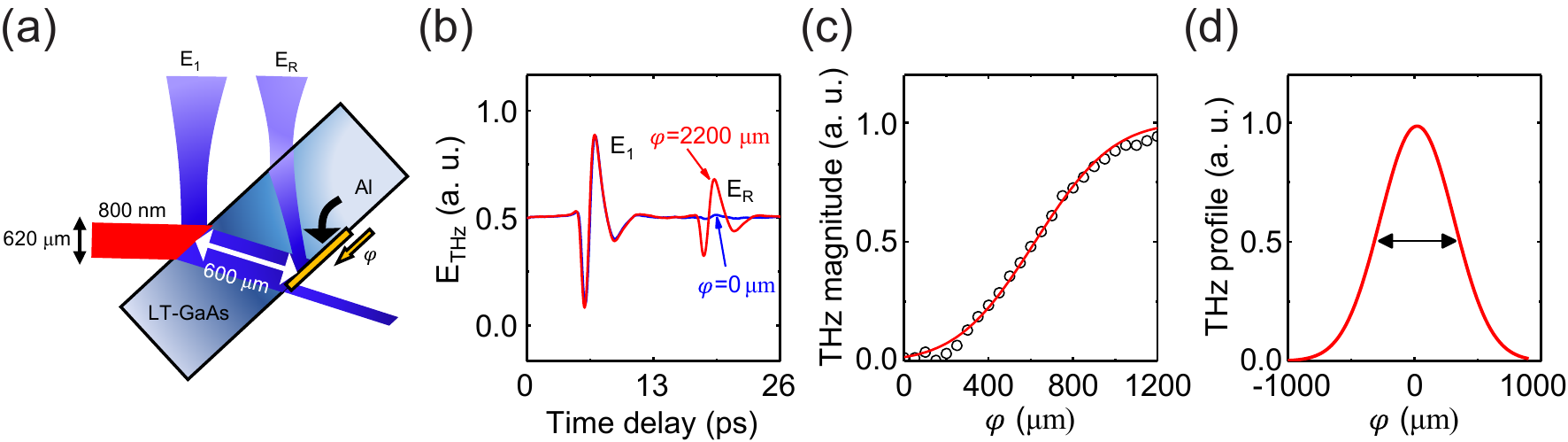} 
\caption{ 
\label{fig:THz2}
Characterization of the THz beam at the reflection spot.
(a) Schematic of the knife-edge experiment to characterize the beam size at the backside of the LT-GaAs THz emitter. 
$\varphi$ denotes a displacement of the knife edge. 
E$_\text{R}$ is the partially reflected THz pulse by an Al-coated region. 
(b) THz electric field recorded in time-domain. Red and blue show THz pulse measured at $\varphi$$=$2200 and 0 $\mu$m, respectively.
(c) Peak-to-peak amplitude of E$_\text{R}$ (symbol) as a function of $\varphi$. 
Gaussian fit to the experimental result is shown with a solid line.
(d) Restored THz profile obtained from the Gaussian fit procedure. 
THz beam width defined in full-width half maximum is about 670 $\mu$m.} 
\end{figure}

To determine the THz beam size at the backside of the LT-GaAs, we carry out the knife-edge experiment. 
We evaporate Al with a rectangular shape onto the backside of the THz emitter which acts as a THz knife-edge reflector, and translate the Al-coated region by $\varphi$ up to 2200 $\mu$m as schematically shown in Fig. \ref{fig:THz2}(a). 
Detailed procedure of the knife-edge experiment is provided in Ref. \cite{han_radiating_2018}. 
In this experiment, we set the beam waist of the pumping laser to 620 $\mu$m, and consider that the radiated THz pulse at the generation point has the same beam waist. 
Figure \ref{fig:THz2}(b) exhibits time-domain THz electric field profiles (E$_1$ and E$_\text{R}$) obtained at $\varphi=0$ and $\varphi=2200$ $\mu$m. 
The stronger pulse corresponds to E$_1$ emitted directly from the emitter surface, 
and the other corresponds to E$_\text{R}$ reflected from the emitter-Al interface. 
As $\varphi$ is varied, E$_\text{1}$ remains the same, but E$_\text{R}$ changes dramatically. 
Normalized THz peak-to-peak amplitude of E$_\text{R}$ against the position $\varphi$ is shown in Fig. \ref{fig:THz2}(c). 
We fit this knife-edge scan result by considering the Gaussian profile of the THz wave, 
and show the fitting curve with a solid line which matches well the experimental result. 
As shown in Fig. \ref{fig:THz2}(d), the Gaussian profile used for this fitting has the FWHM of about 670 $\mu$m which is close to the original THz beam size at the generation point as expected. 
Although the THz beam has a spectral content between 0.3 and 1.2 THz, the beam waist is given with little dependence on the frequency since the Rayleigh length is long enough compared to the emitter thickness, namely 1.2 mm at 0.5 THz. Consequently, we can perform the absolute reflectivity measurement on the 800 $\mu$m wide sample even in the THz spectral region by having the THz beam size smaller or comparable to the sample size.

\subsection{Reliability of extracted optical conductivity spectra}

\begin{figure}[h]
\includegraphics[width=160mm]{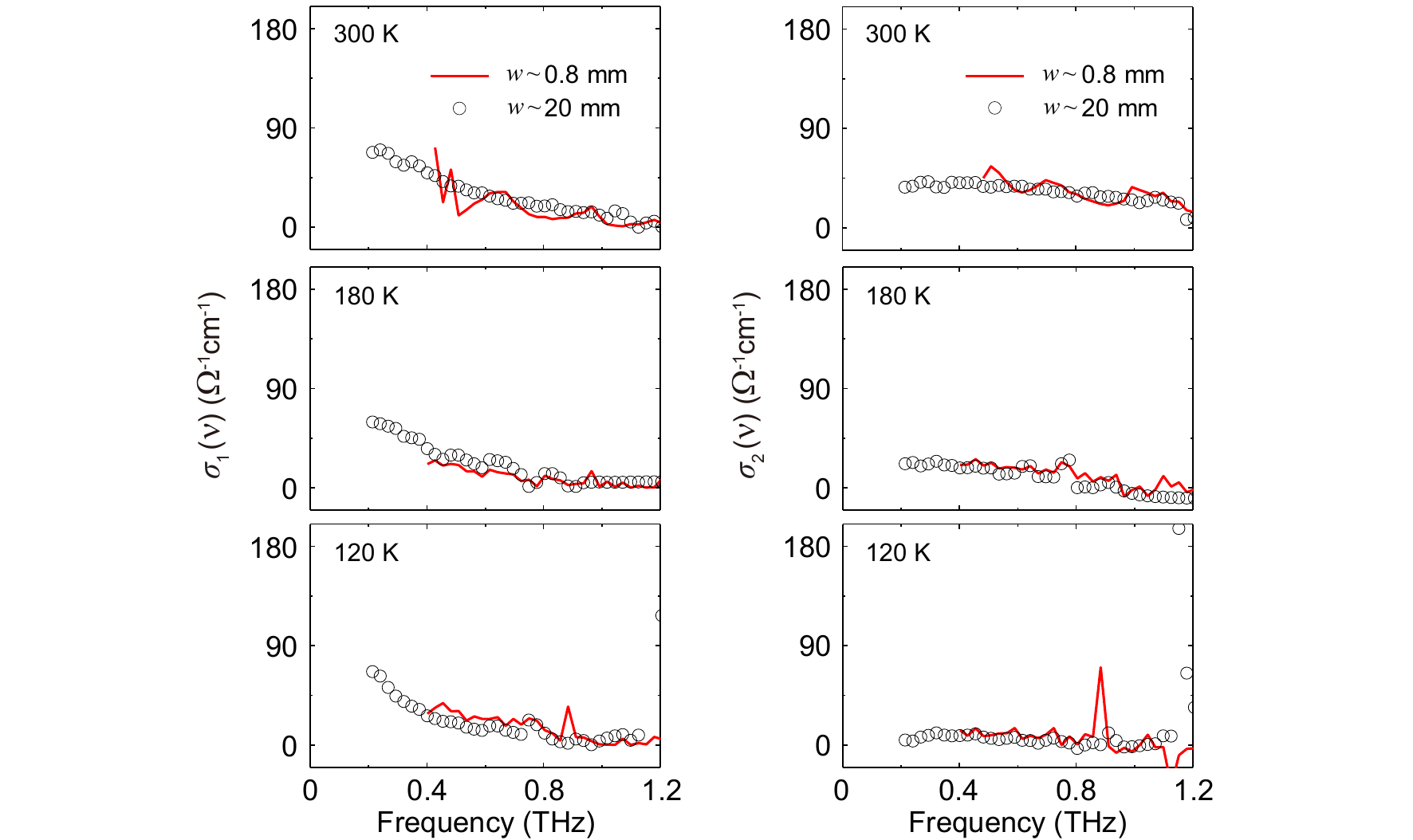} 
\caption{ 
\label{fig:THz3}
Complex optical conductivity spectra of n-InAs at 300, 180, and 120 K.
Open symbols and lines are results for the samples of different sizes, i.e., $\textit{w}=0.8$ and $20$ mm, respectively. } 
\end{figure}

To confirm that this approach is really applicable to determine optical conductivity spectra of the sample of which size is the comparable to the wavelength of THz light, 
we test n-InAs having the size of $0.8\times0.8$ mm$^2$ ($\textit{w}=0.8$ mm) with a rectangular shape. 
For the comparison, we prepare the large-sized, i.e., 20 mm wide sample. 
Figure \ref{fig:THz3} displays real and imaginary parts of optical conductivity spectra obtained at 120, 180, and 300 K. 
Open symbols and lines are the results for $\textit{w}=0.8$ mm and $\textit{w}=20$ mm, respectively, 
and they match quite well. 
The frequency region lower than 0.4 THz is not accessible for the small sample due to the diffraction-limit. 
This confirms that we are able to obtain optical constants of the small-sized sample less than 1 mm even with a variation of $T$ in the frequency range from 0.4 THz to 1.2 THz.

\newpage

\section{\normalsize E\lowercase{xtraction of} $D$ \lowercase{and} $\bold{\gamma}$ \lowercase{from the} D\lowercase{rude-}L\lowercase{orentz fitting}}

\subsection{Drude-Lorentz fitting of the optical conductivity spectra}

We here explain the fitting procedure of the optical conductivity spectra $\tilde{\sigma}$ using the Drude and the Lorentz models which account for optical response from free carriers and bound charges, respectively. 
The total $\tilde{\sigma}$ is given by $\tilde{\sigma} = \sigma_1(\nu) + i\sigma_2(\nu) = \frac{D}{2}\frac{1}{\gamma-i\nu} + \sum_j \frac{L_j}{2} \frac{\nu}{i(\nu_j^2-\nu^2)+\nu \Gamma_j}$. 
In the Drude response appearing in the first part, $D$ and $\gamma$ are the Drude weight and the scattering rate of free carriers, respectively. 
In the Lorentz oscillator response appearing in the second part, $L_j$, $\Gamma_j$ and $\nu_j$ are the strength, damping rate, and center frequency of the $j$-th oscillation, respectively.
We determine $D$ and $\gamma$ by fitting $\tilde{\sigma}$ by means of the nonlinear regression method. 
Importantly, a satisfactory fit should give the dc-limit conductivity consistently with the actual dc-conductivity obtained independently.
Details of optical contribution from bound charges are provided below.

\subsection{Consideration of phonon and interband transition}

\begin{figure}[h]
\includegraphics[width=\textwidth]{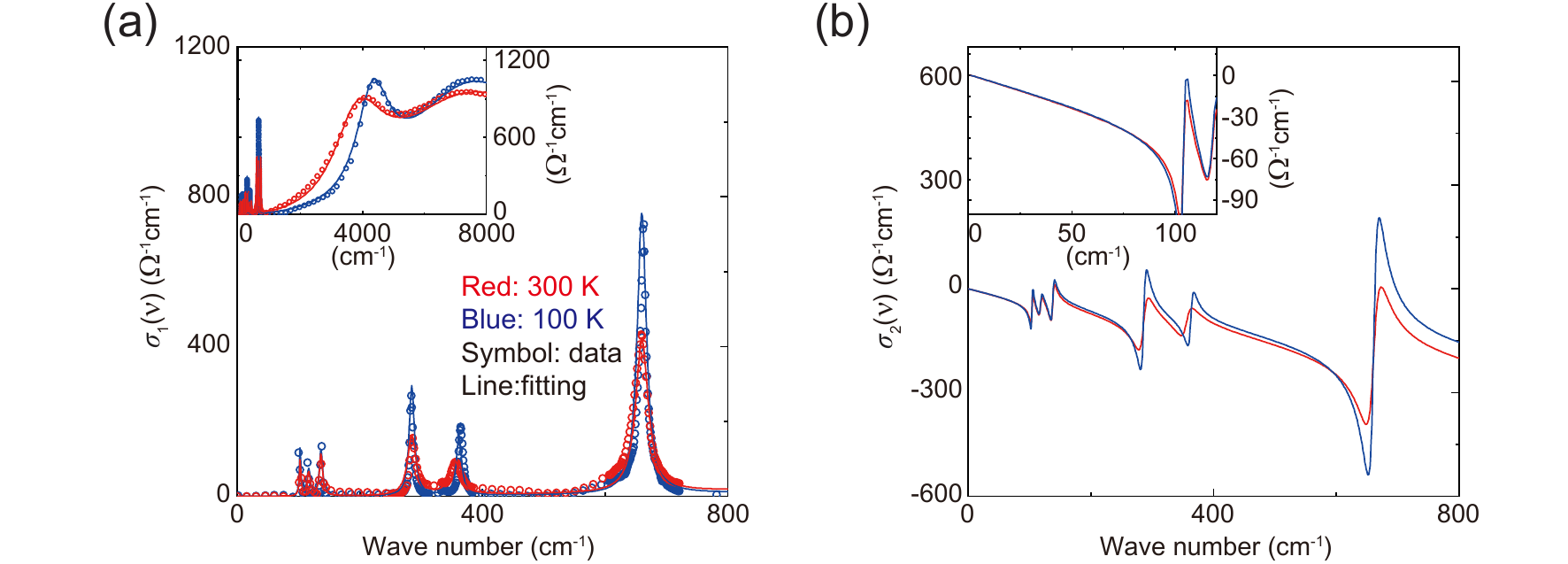} 
\caption{ 
\label{fig:THz4}
(a) Real part $\sigma_1(\nu)$ and (b) imaginary part $\sigma_2(\nu)$ of optical conductivity spectra of Sr$_2$IrO$_4$ at 100 and 300 K.
The result of $\sigma_1(\nu)$ is taken from Ref. \cite{moon_temperature_2009}.
In (a), open circles are experimental data, and solid lines show the Lorentz fit results.
Sharp six peaks correspond to optical phonons. 
Inset shows $\sigma_1(\nu)$ in a broader frequency range below 8000 cm$^{-1}$.
In (b), $\sigma_2(\nu)$ obtained from Lorentz fitting result of (a). 
Inset shows $\sigma_2(\nu)$ in a lower frequency range below 120 cm$^{-1}$. } 
\end{figure}

We assume phonon and interband transition would not vary significantly upon the 3\% Tb doping in Sr$_2$IrO$_4$ 
and take the results of Sr$_2$IrO$_4$ from Ref. \cite{moon_temperature_2009} for considering the high-energy contribution. 
Figure \ref{fig:THz4}(a) shows $\sigma_1(\nu)$ of Sr$_2$IrO$_4$ at 100 K and 300 K, which highlight six optical phonon responses. 
$\sigma_1(\nu)$ in a broad frequency range is provided in its inset where interband transitions are observed. 
Lorentz model is employed to reproduce the six optical phonons below 800 cm$^{-1}$ as well as the interband transitions located at about 4000 and 8000 cm$^{-1}$. 
The higher energy contributions are considered as $\epsilon_{\infty} \sim 6$. 
As can be seen, Lorentz model fits $\sigma_1(\nu)$ at both 100 K and 300 K well. 
From this fitting procedure, we ascertain the imaginary part of optical conductivity $\sigma_2(\nu)$ shown in Fig. \ref{fig:THz4}(b). 
The inset of Fig. \ref{fig:THz4}(b) exhibits $\sigma_2(\nu)$ in the low frequency part below 120 cm$^{-1}$. 
As $T$ is varied from 100 K to 300 K, 
widths as well as peak positions of phonons and interband transitions exhibit noticeable changes, 
but it is important to note that the optical conductivity spectrum below 50 cm$^{-1}$ remains almost the same. 
Therefore, we treat $\sigma_1(\nu)$ and $\sigma_2(\nu)$ shown in Fig. \ref{fig:THz4} as the Lorentz oscillator contribution in fitting $\tilde{\sigma}(\nu)$ of Tb-doped Sr$_2$IrO$_4$. 

\subsection{Drude-Lorentz fitting results}
 
\begin{figure}[b]
\includegraphics[width=\textwidth]{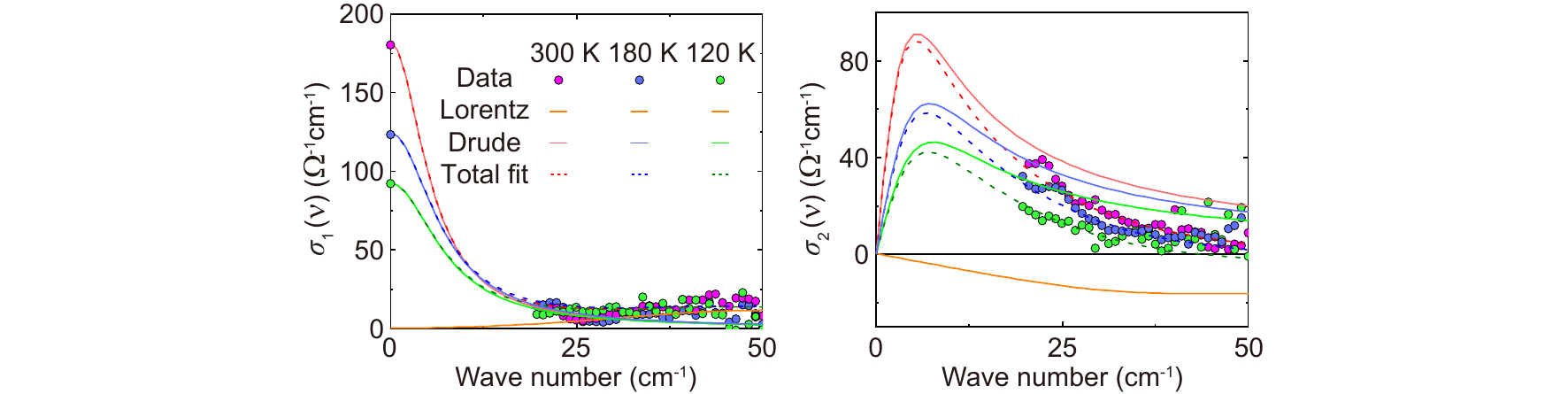} 
\caption{ 
\label{fig:THz5}
Drude-Lorentz fitting results for optical conductivity spectra of Sr$_2$(Ir$_{0.97}$Tb$_{0.03})$O$_4$ at selected temperatures of 300, 180, and 120 K.   
} 
\end{figure}

Figure \ref{fig:THz5} displays the Drude-Lorentz fitting results for optical conductivity spectra at 300, 180, and 120 K. 
These data sets are the same with those of Fig. 3(c) in the main text. 
For each data set, Drude and Lorentz contributions are shown with solid lines and the total fit is denoted by a dotted line.
The low-frequency tail of Lorentz contribution presented in Fig. \ref{fig:THz4}(a) can account for the finite values of experimental $\sigma_1(\nu \gtrsim 30~\text{cm}^{-1})$.  
Whereas the Lorentz contribution has a large portion for $\sigma_2(\nu)$, its temperature-dependence is ignorable as presented in Fig. \ref{fig:THz4}(b), which allows us to relate the temperature-dependent changes of $\sigma_2(\nu)$ to the temperature-dependent changes of Drude response.
 
\begin{figure}[h]
\includegraphics[width=\textwidth]{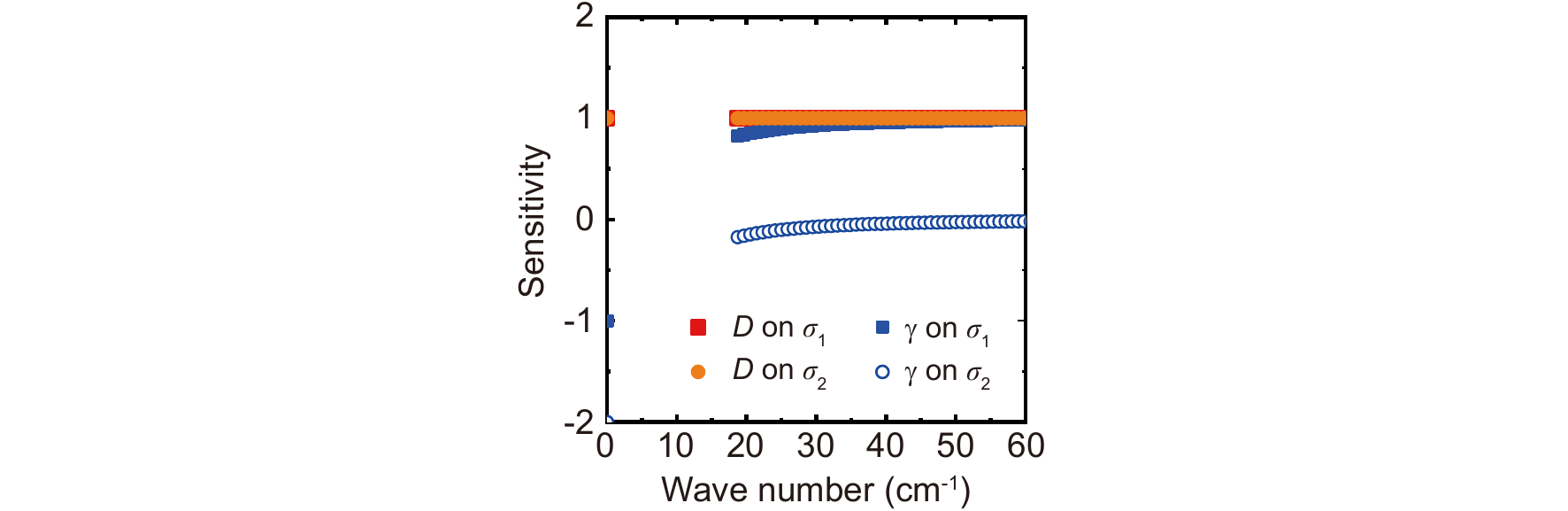} 
\caption{ 
\label{fig:THz6}
Sensitivities of Drude fit process to the Drude weight $D$ and the scattering rate $\gamma$. Isolated lowest frequency values are ones in the zero-frequency limit for each case. 
} 
\end{figure}
 
Whether the Drude analysis is valid or not depends on the sensitivity of the optical conductivity spectra to the Drude fit parameters. 
The sensitivity coefficient is defined as 
$S_x (\sigma_{1,2} ) = \frac{dln(\sigma_{1,2})}{dln(x)} =\frac{x}{\sigma_{1,2}}\frac{d\sigma_{1,2}}{dx}$, where $x=D$ or $\gamma$. 
Figure \ref{fig:THz6} displays sensitivities of each Drude parameter for the real and imaginary parts of optical conductivity, showing that the Drude fit analysis has high enough sensitivities to both $D$ and $\gamma$. 
The sensitivity to $D$ is given as the unity which simply follows the definition of the sensitivity and the Drude formula; 10 \% increase in $D$ will result in 10 \% increase in both $\sigma_1$ and $\sigma_2$. 
We checked also the correlation of two fitting parameters which is estimated to be about 0.66. 
Although this value indicates a relatively large correlation between $D$ and $\gamma$, such high sensitivities to $D$ and $\gamma$ guarantee the satisfactory Drude fit which should allow the reliable determination of $D$ and $\gamma$ using the given experimental data with a limited spectral range. 
Actually, we artificially generated data points in the originally missing spectral range to have the same fluctuation level with original data above 20 cm$^{-1}$, and made the same Drude fit analysis. 
Importantly, the obtained parameters are essentially the same with those obtained for the original experimental data whereas the error becomes reduced simply due to the wider spectral range of valid data points. 
This ascertains that available experimental results contain enough information which can lead to the unambiguous determination of Drude parameters. 
 
Although we assumed that the THz results available down to about 20 cm$^{-1}$ should be connected to the dc-limit values by following the simple Drude model, it is worth to check other possibilities of the spectral behaviors in the missing spectral region. 
Among various possibilities, we can easily exclude any contribution from the bound excitations. 
Interband transitions and optical phonons should appear in the much higher photon energy range as demonstrated in the Section 6.2. 
Considering spectral weight and line width observed in our THz data at 300 K, we can exclude also the hydrogen-like impurity excitations and magnon contributions which should appear with a much smaller linewidth at the very low temperature and in the magnetically ordered state, respectively. 
By the way, it would seem feasible to consider the so-called localization modified Drude model which can account for the possible disorder effect arising from the Tb ion substituted in Sr$_2$IrO$_4$ ~\cite{Lee_reflectance_1993,Tzamalis_doping_2002,Kim_metal_2005}.
The disorder effect typically appears as a peak structure in the $\sigma_1$ spectrum at the finite frequency, and the frequency is in proportion to the disorder effect. 
If the $\sigma_1$ spectrum were to have a disorder-induced peak at the finite frequency, the peak frequency should be located at the lower frequency than 20 cm$^{-1}$ which is a much smaller value when it is compared with other disordered systems having similar amount of doping concentrations ~\cite{Cooke_electron_2006,Hempel_intragrain_2016}.
This implies that the disorder effect should be small even if it would exist in Sr$_2$(Ir$_{0.97}$Tb$_{0.03})$O$_4$. 
In particular, the fit to $\sigma_1$ and $\sigma_2$ based on the model results in the scattering rate which is an order of magnitude smaller than what we obtained from the simple Drude model. 
We therefore do not consider such localization-induced modification in the Drude model, and instead take the simple Drude model to account for the experimental results. 

\subsection{Temperature-dependent variation of the Drude weight}

\begin{figure}[h]
\includegraphics[width=\textwidth]{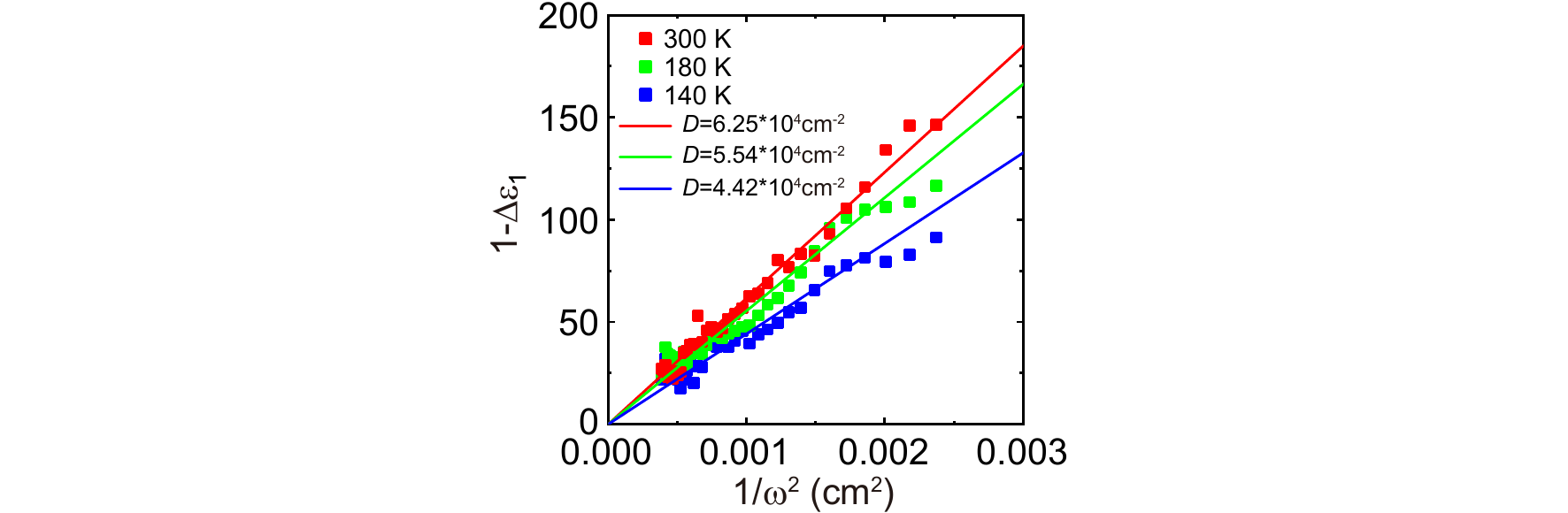} 
\caption{ 
\label{fig:THz7}
Spectral behaviors of the real part of the dielectric constant in the high frequency side.
} 
\end{figure}

We take into account the temperature-dependent variation of the Drude weight as a key experimental evidence for supporting the Dirac point node rather than the Dirac line node in Tb-doped Sr$_2$IrO$_4$. 
To confirm the validity of this argument, we double check the Drude weight by examining the real part of a dielectric constant spectrum $\varepsilon_1(=1-4\pi\sigma_2/\omega)$.
According to Drude formalism, $\varepsilon_1$ in the high frequency region behaves as $\Delta\varepsilon_1\cong 1-D/\omega^2$, where $\Delta\varepsilon_1$ denotes the free carrier contribution. 
Figure \ref{fig:THz7} shows that $1-\Delta\varepsilon_1$ scales well with $1/\omega^2$,
implying that the high frequency response accords with the Drude prediction. 
We plot also straight lines with proportional coefficients taken from the Drude weight shown in Fig. 4(b), 
which are determined by the Drude fit at each temperature, and find that they match well the experimental data of $1-\Delta\varepsilon_1$.
This demonstrates that our analyses based on the simple Drude model are fully self-consistent in determining the temperature-dependent Drude weight. 
 
\subsection{Error estimation for the Drude-Lorentz fitting}

\begin{figure}[h]
\includegraphics[width=\textwidth]{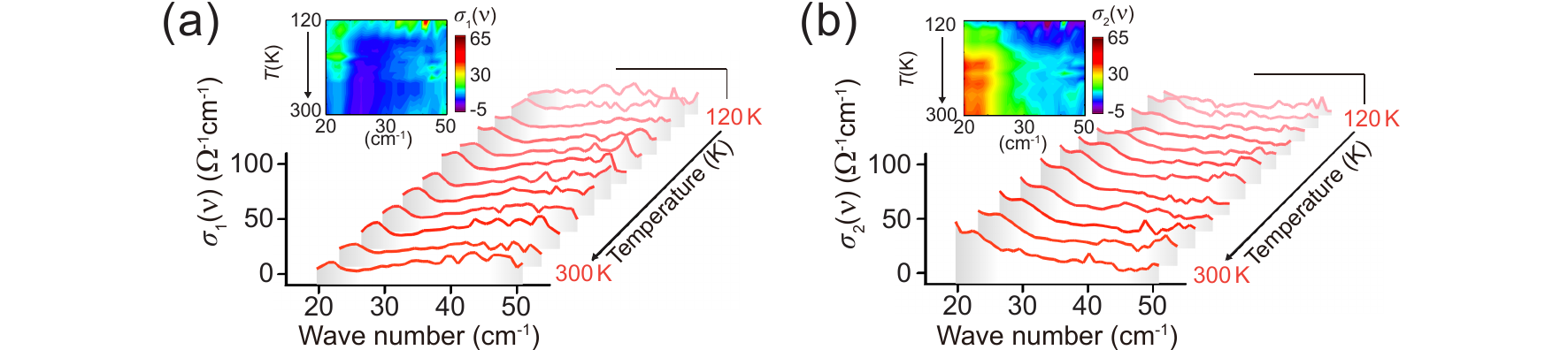} 
\caption{ 
\label{fig:THz8}
Temperature-dependent optical conductivity spectra. (a) Real and (b) imaginary parts of optical conductivity spectra from 120 K to 300 K. The inset of each figure displays the 2D contour map of the optical conductivity with variations of temperature and wavenumber.  
} 
\end{figure}

As can be seen in Fig. 3 in the main text and also in Fig. \ref{fig:THz8}, 
optical conductivity data have point-to-point fluctuations, 
and there should be finite fitting errors originating from such fluctuations in experimental data. 
In the course of the Drude fit using the nonlinear regression method, we set the confidence interval as 99.8 \% to estimate the extent of errors. 
Note that we computatively confirmed that 99.8 \% is the maximum tolerance to give the largest error bars in the nonlinear regression method we adopted for the error estimation.
Hump and spike structures observed in optical conductivity spectra are likely due to experimental artifact so that we here refrain from assigning any physical meaning to those features.
Corresponding errors are included in plotting the best fit results of $D$ and $\gamma$ in Fig. 4, manifesting that the characteristic $T$-dependent tendencies of $D$ and $\gamma$ can be clearly caught far beyond the error bars. 

\begin{figure}[h]
\includegraphics[width=\textwidth]{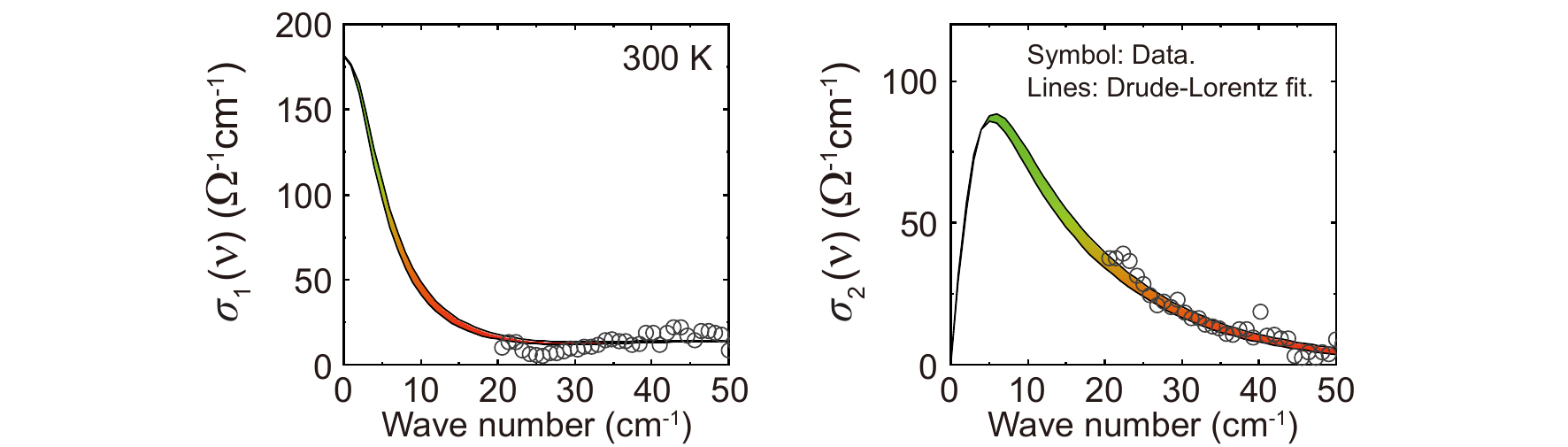} 
\caption{ 
\label{fig:THz9}
Results of the Drude-Lorentz fit for optical conductivity spectra of Sr$_2$(Ir$_{0.97}$Tb$_{0.03})$O$_4$ at 300 K.
Confidence bands corresponding to the 99.8 \% confidance interval are visualized using the filling curve.  
} 
\end{figure}

In Fig. \ref{fig:THz9}, we visualize the 99.8 \% confidence bands using filling curves which could cover most of data points in the $\sigma_2$ spectrum. 
Meanwhile, there seems a relatively large deviation in the $\sigma_1$ spectrum. 
This is attributed to the fact that the Drude response is narrowly confined and has a minimal contribution to $\sigma_1(\nu)$ above 20 cm$^{-1}$. 
In other words, such deviations stem from the uncertainty in the high energy contributions. 

\clearpage

\section{\normalsize B\lowercase{oltzmann transport theory for optical} \lowercase{properties of the} DPN \lowercase{state}}

In this section, we calculate the intraband optical conductivity by using semi-classical Boltzmann transport equation.
First, we discuss the Boltzmann transport theory for anisotropic system
because DPN has the anisotropic electronic band structure.
Then we present the details of phenomenological model for the scattering rate 
and calculate the scattering rate using random phase approximation (RPA) for effective Hamiltonian in Section 2.
After that, we compare the RPA results with the phenomenological model and confirm the validity of the phenomenological model.
Finally, we compare the temperature-dependent optical properties of DPN state with experimental data.
%
%
%

\subsection{Boltzmann transport theory for anisotropic systems}
The linearized time-independent Boltzmann transport equation for the distribution function $f({\mathbf k}) = f^{(0)}(\mathbf k) + \delta f ({\mathbf k})$ is given by
\begin{eqnarray} \label{seq:Boltzmann_Eq_01}
(-e) \mathbf E \cdot \mathbf {v}_{\mathbf k} S^{(0)}(\epsilon)
= \int \frac{d^2 k\rq{}}{ (2\pi)^2} W_{\mathbf{k}\mathbf{k}\rq{}} (\delta f(\mathbf k)-\delta f(\mathbf k\rq{}) ),
\end{eqnarray}
where $f^{(0)}(\mathbf k) = ( \exp [\beta( \epsilon_{\mathbf k} - \mu)] +1 )^{-1}$ is the Fermi-Dirac distribution at equilibrium with $\beta=1/k_BT$,
$\delta f ({\mathbf k}) $ is the deviation proportional to applied electric field $\mathbf E$, $\mathbf {v}_{\mathbf k}$ is the velocity of electron, 
$S^{(0)}(\epsilon) = - \frac{ \partial f^{(0)}(\epsilon)}{\partial \epsilon}$,
$ W_{ \mathbf{k}\mathbf{k}' } = \frac{2\pi}{\hbar}n_{\text{imp}}|V_{ \mathbf{k}\mathbf{k}' }| ^2 \delta(\epsilon_{\mathbf{k}}-\epsilon_{\mathbf{k}'})$,
$n_{\text{imp}}$ is the impurity density
and $V_{ \mathbf{k}\mathbf{k}' }$ is the matrix element of the scattering potential.
For isotropic systems, the relaxation time $\tau_{\mathbf k}$ within the relaxation time approximation is given by
\begin{eqnarray}\label{seq:relaxation_time_integral_iso}
\frac{1}{\tau_{\mathbf k}} 
=
\int d^2 k \rq{}
W_{ \mathbf{k}\mathbf{k}\rq{} }
\left (1 -  \cos \theta_{\mathbf k \mathbf k\rq{}} \right ),
\end{eqnarray}
where 
$
\theta_{\mathbf k \mathbf k\rq{}}
$
is the scattering angle between momenta $\mathbf k$ and $\mathbf k\rq{}$.

For an anisotropic system, the relaxation time depends on the  momentum direction and hence equation (\ref{seq:relaxation_time_integral_iso}) for isotropic system needs to be modified.
Up to linear order of external electric field $\mathbf E$,
we use the following ansatz for $\delta f ({\mathbf k})$:
\begin{eqnarray}\label{seq:ansatz_02}
\delta f ({\mathbf k}) 
= (-e) \left(   \sum_{i=1}^{2} E^{(i)} {v}_{\mathbf k}^{(i)} {\tau}_{\mathbf k}^{(i)}    \right ) 
S^{(0)}(\epsilon),
\end{eqnarray}
where $ E^{(i)}, {v}_{\mathbf k}^{(i)} $, and 
$ {\tau}_{\mathbf k}^{(i)}$ ($i=x,y$) are the $i$th component of electric field, velocity, and relaxation time, respectively.
Then, by inserting equation (\ref{seq:ansatz_02}) into equation (\ref{seq:Boltzmann_Eq_01}) and matching each coefficient of $ E^{(i)} $, we get the following integral equation for the relaxation time:
\begin{eqnarray} \label{seq:relaxation_anisotropic}
1 = \int \frac{d^2 k\rq{}}{ (2\pi)^2} W_{\mathbf{k}\mathbf{k}\rq{}}
\left(   \tau_{\mathbf k}^{(i)} - \frac{v_{\mathbf k\rq{}}^{(i)}}{v_{\mathbf k}^{(i)}}  \tau_{\mathbf k\rq{}}^{(i)}  \right ).
\end{eqnarray}
This coupled integral equation can be solved by the numerical method \cite{liu_mobility_2016,park_semiclassical_2017}.

Now, we calculate the optical conductivity induced by the externally applied time-dependent electric field $\mathbf E(t)$.
The linearized time-dependent Boltzmann equation along $i$th direction in the relaxation time approximation is given by
\begin{eqnarray}
\frac{ \partial \delta f(\mathbf k, t)}{ \partial t} + e E^{(i)}(t) {v}_{\mathbf k}^{(i)}
S^{(0)}(\epsilon)
& = &
- \frac{ \delta f(\mathbf k, t)}{ \tau^{(i)}_{\mathbf k}}.
\end{eqnarray}
Using the Fourier transformations $E^{(i)}(t) = \int E^{(i)}(\omega) e^{- i \omega t} d\omega$ and $\delta f(\mathbf k, t) = \int  \delta f(\mathbf k,\omega) e^{- i \omega t} d\omega$,
we find a solution for $\delta f$ along $i$th direction as
\begin{eqnarray}
\delta f(\mathbf k, \omega) & = & 
\frac{(-e) E^{(i)}(\omega) {v}_{\mathbf k}^{(i)}}{ (\tau^{(i)}_{\mathbf k})^{-1} - i \omega }
S^{(0)}(\epsilon).
\end{eqnarray}
Then the current density $J^{(i)}(\omega)$ is given by
\begin{eqnarray}
J^{(i)}(\omega)  =  
g (-e)
\int \frac{d^2 k}{(2 \pi)^2} 
{v}_{\mathbf k}^{(i)} \delta f(\mathbf k, \omega)
  =  g e^2
\int \frac{d^2 k}{(2 \pi)^2} 
\frac{{v}_{\mathbf k}^{(i)} {v}_{\mathbf k}^{(j)} E^{(j)}(\omega)}{ (\tau^{(j)}_{\mathbf k})^{-1} - i \omega }
S^{(0)}(\epsilon),
\end{eqnarray}
where $g$ is the degeneracy factor. 
Since $J^{(i)}(\omega) = \sigma_{ij}(\omega)E^{(j)}(\omega)$, the optical conductivity $\sigma_{ij}(\omega)$ is given by 
\begin{eqnarray}\label{seq:optical_cond}
\sigma_{ij} (\omega)  & = & 
g e^2
\int \frac{d^2 k}{(2 \pi)^2} 
\frac{ {v}_{\mathbf k}^{(i)} {v}_{\mathbf k}^{(j)} }{ \gamma^{(j)}_{\mathbf k}- i \omega }
S^{(0)}(\epsilon),
\end{eqnarray}
where $ \gamma^{(i)}_{\mathbf k} = 1/\tau^{(i)}_{\mathbf k} $ is the momentum-dependent scattering rate along the $i$th direction.
Note that DC conductivity $\sigma_{\text{DC}}^{ij}=\sigma_{ij}(\omega=0)$.

\subsection{Phenomenological model for the scattering rate}

The phenomenological model is constructed by the linear combination of short-and long-range charged impurities according to the Matthiessen's rule.
Since we consider charged impurity scattering, $W_{ \mathbf{k}\mathbf{k}' }$ becomes
\begin{eqnarray}
W_{ \mathbf{k}\mathbf{k}' }
= \frac{ 2 \pi n_{\text{imp}} }{ \hbar }  \abs{ U (q) }^2   \left(\frac{1 + \cos \theta_{\mathbf k \mathbf k\rq{}}}{2}\right) \delta(E_{\mathbf k}-E_{\mathbf k\rq{}}),
\end{eqnarray}
where $q=|\mathbf{k}-\mathbf{k}'|$ and $ U (q)$ is the screened Coulomb potential of charged impurity.
For a short-range charged impurity, the screened Coulomb potential $U_{\text{short}}(q)$ becomes completely screened and hence becomes a constant as $U_{\text{short}}(q)=U_{\text{short}}$.
Meanwhile, for a long-range charged impurity, the screened Coulomb potential $U_{\text{long}}(q)$ becomes unscreened as $U_{\text{long}}(q)=(2\pi e^2)/\kappa q$
where $\kappa$ is the effective dielectric constant.
Given this, we obtain the energy-dependent scattering rates for short-range and long-range charged impurities as
\begin{eqnarray}
\gamma_{\text{short}}  (E_{\mathbf k})
=
\gamma^{(0)}_{\text{short}} \left( \frac{ E_{\mathbf k} } { E_\text{F} } \right)
~~
\text{and}
~~
\gamma_{\text{long}}  (E_{\mathbf k})
=
\gamma^{0}_{\text{long}} \left( \frac{ E_\text{F} } { E_{\mathbf k} } \right),
\end{eqnarray}
where $\gamma^{0}_{\text{short}}$ and $\gamma^{0}_{\text{long}} $ are characteristic scattering rates at the reference temperature. 
Here, for the DPN and DLN states, we numerically calculated these energy-dependent scattering rate using the effecitve Hamiltonian including the dispersion of DPN and DLN. 
These energy-dependent scattering rates were also obtained for graphene in Ref. \cite{hwang_screening-induced_2009_2}. 
With these scattering rates, we construct the phenomenological model for scattering rate as 
\begin{eqnarray}
\gamma_{\text{model}}  (T,E)
=
W_1(T) \gamma_{\text{short}}(E) + W_2(T) \gamma_{\text{long}}(E),
\end{eqnarray}
where $W_1(T)$ and $W_2(T)$ are adjusted $T$-dependent weights which reproduce the experimentally observed scattering rate in Fig. 4(a).
In the main text, using $\gamma_{\text{model}}  (T,E)$ as input, 
we calculated optical conductivity $\sigma_{\text{model}}  (\nu)$ for the calculated band structures in Figs. 2(a) and 2(b); 
then obtained $\gamma(T)$, $D(T)$ and $\sigma_{\text{DC}} (T)$ for both DPN and DLN from a Drude fit of $\sigma_{\text{model}}  (\nu)$ [Figs. 4(d)-(f)].

\subsection{RPA scattering rate}

Now, we calculate anisotropic scattering rate using RPA for constructed effective Hamiltonian.
Within RPA, $U(q)$ is given by
\begin{eqnarray}\label{seq:pol_U}
U(q,T)
= \frac{v_c(q)}{\epsilon(q,T)}
= \frac{2\pi e^2}{\kappa q+2\pi e^2 \Pi(q,T)},
\end{eqnarray}
where $\Pi(q,T)$ is the RPA polarizability function.
The RPA polarizability function $\Pi(q,T)$ is given by the bare bubble diagram \cite{hwang_screening-induced_2009_2}
\begin{eqnarray}
\Pi(q,T)
=-\frac{g}{A}\sum_{\mathbf k,s,s'}\frac{f_{s \mathbf k}-f_{s' \mathbf{k'}}}{E_{s \mathbf k}-E_{s' \mathbf{k'}}}\frac{1+ss' \cos\theta_{ \mathbf k \mathbf{k'}}}{2},
\end{eqnarray}
where $A$ is the area of the system and $f_{s \mathbf k} = (\exp[\beta(E_{s \mathbf k}-\mu)]+1)^{-1}$ with $\beta=1/k_BT$.
Using calculated $\Pi(q,T)$ in Fig. 4(g), we calculate the angle-dependent scattering rate (equation (\ref{seq:relaxation_anisotropic})) in the long-wavelength limit $q\rightarrow 0$ (Thomas-Fermi approximation) and
average over the angle dependence of calculated scattering rate as shown in Fig. 4(i).

\subsection{Comparison between phenomenological model and RPA}

\begin{figure}[h]
\includegraphics[width=\textwidth]{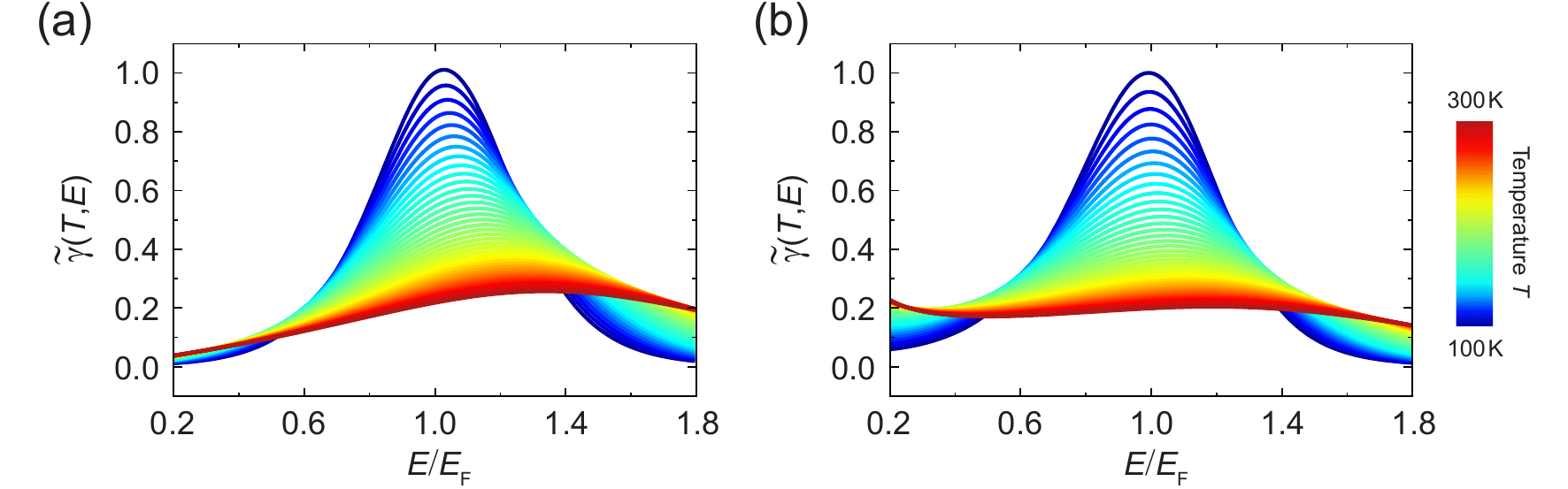} 
\caption{ 
\label{fig:Mixed_pol}
(a),(b) Normalized scattering rate $\tilde{\gamma}(T,E)$ of (a) the anisotropy considered RPA and (b) the phenomenological model.
} 
\end{figure}

In the main text, we used the phenomenological model for scattering rate to compare between theory and experiment (Fig. 4). 
We now confirm its validity.
To compare scattering rates between RPA result and phenomenological model, we define a normalized scattering rate $\tilde{\gamma}(T,E)$ as
\begin{eqnarray}
\tilde{\gamma}(T,E)
= \frac{\gamma(T,E) (-\frac{\partial f(T,E)}{\partial E})}{\gamma(100 \text{K},E_{\text{F}}) (-\frac{\partial f(100 \text{K},E)}{\partial E} \big\rvert_{E=E_{\text{F}}})},
\end{eqnarray}
which plays a dominant role in the calculation of optical conductivity [equation (\ref{seq:optical_cond})].
Figure \ref{fig:Mixed_pol} shows very similar temperature and energy depedences between two normalized scattering rates.
Thus, we can conclude that our adjusted weights $W_1(T)$ and $W_2(T)$ in the phenomenological model well reproduces the RPA screening effect.

\subsection{Optical properties of anisotropic DPN state}

\begin{figure}[h]
\includegraphics[width=\textwidth]{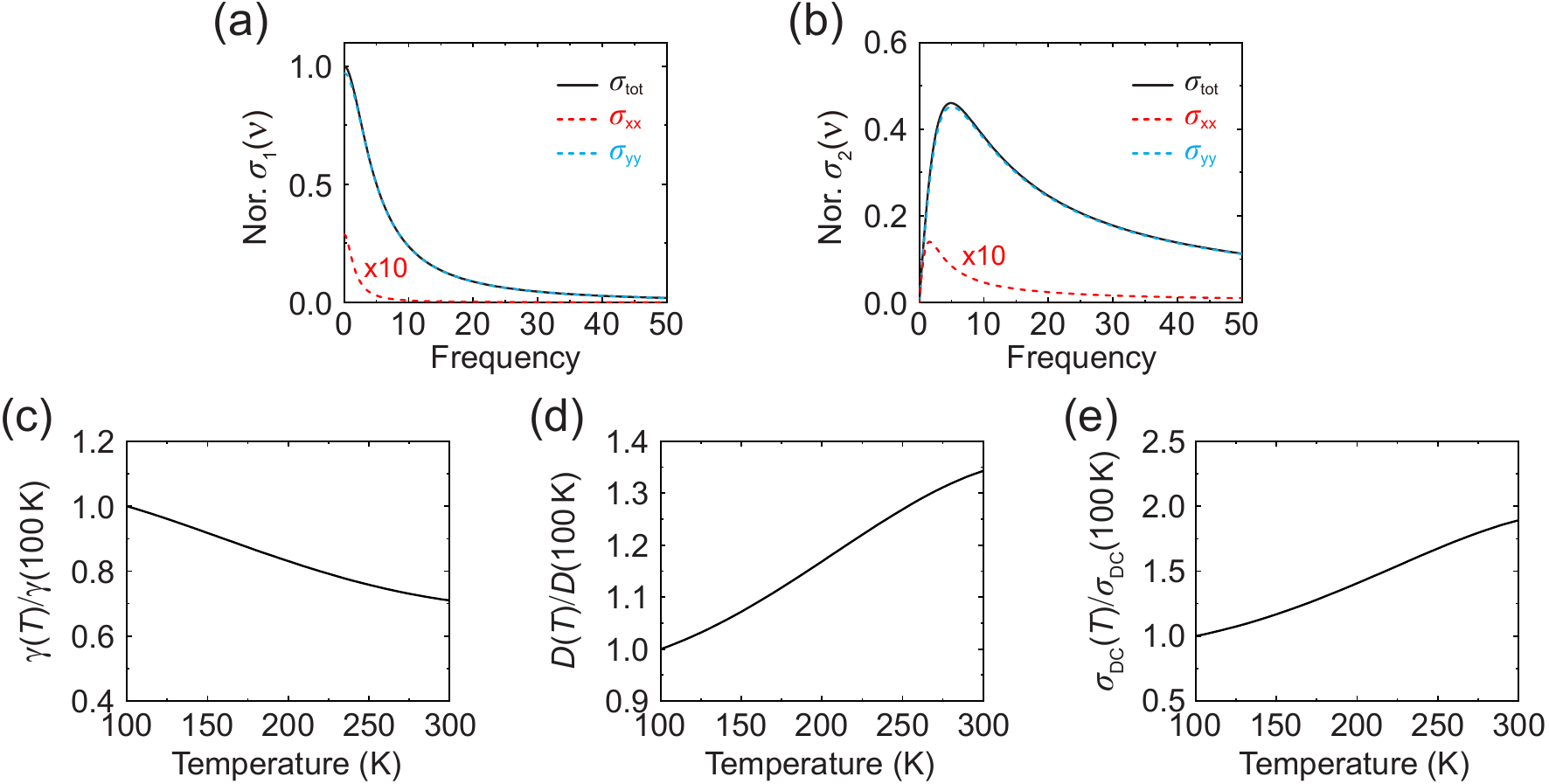}
\centering
\caption{ 
\label{fig:eff_Ham_Tb}
(a),(b) Direction-dependent real $\sigma_1(\nu)$ and imaginary $\sigma_2(\nu)$ parts of optical conductivity calculated for the effective Hamiltonian near the Y point. 
Here, $\sigma_1(\nu)$ and $\sigma_2(\nu)$ are normalized by the total value of $\sigma_1 (\nu=0)$.
(c)-(e) Normalized temperature-dependent scattering rate $\gamma(T)$, Drude weight $D(T)$, and DC conductivity $\sigma_{\text{DC}}(T)$.
} 
\end{figure}

Now, we calculate temperature-dependent optical conductivity (equation (\ref{seq:optical_cond})) for effective Hamiltonian using RPA scattering rate [Fig. 4(i)] as input.
Since the Dirac point is anisotropic, the calculated optical conductivity shows anisotropic behaviors along $x$ and $y$ directions [Figs. \ref{fig:eff_Ham_Tb}(a) and \ref{fig:eff_Ham_Tb}(b)];
the calculated optical conductivity $\sigma_{yy}$ is larger than $\sigma_{xx}$ for effective Hamiltonian near the Y point.
Because there are two distinct Dirac points at the X and Y points, the total optical conductivity is the sum of $\sigma_{xx}$ and $\sigma_{yy}$.
We have confirmed that the calculated optical conducitvities $\sigma_{xx}$ and $\sigma_{yy}$ show the typical Drude response  from 100 K to 300 K. 
Then, by fitting the calculated optical conductivity to Drude model $\frac{D}{\gamma-i\omega}$, we extract the temperature-dependent scattering rate $\gamma(T)$, Drude weight $D(T)$ and DC conductivity $\sigma_{\text{DC}}(T)$ in Figs. \ref{fig:eff_Ham_Tb}(c)-(e).
We find that the temperature depedences of $\gamma(T)$, $D(T)$ and $\sigma_{\text{DC}}(T)$ are in very good agreement with experimental data in Fig. 4.
Note that, compared to the results of phenomenolgoical model [Figs. 4(d)-(f)], the anisotropy effect further increases Drude weight $D(T)$, which leads to better agreement with experimental data.

\section{\normalsize C\lowercase{omparison with graphene}}

\begin{figure}[h]
\includegraphics[width=\textwidth]{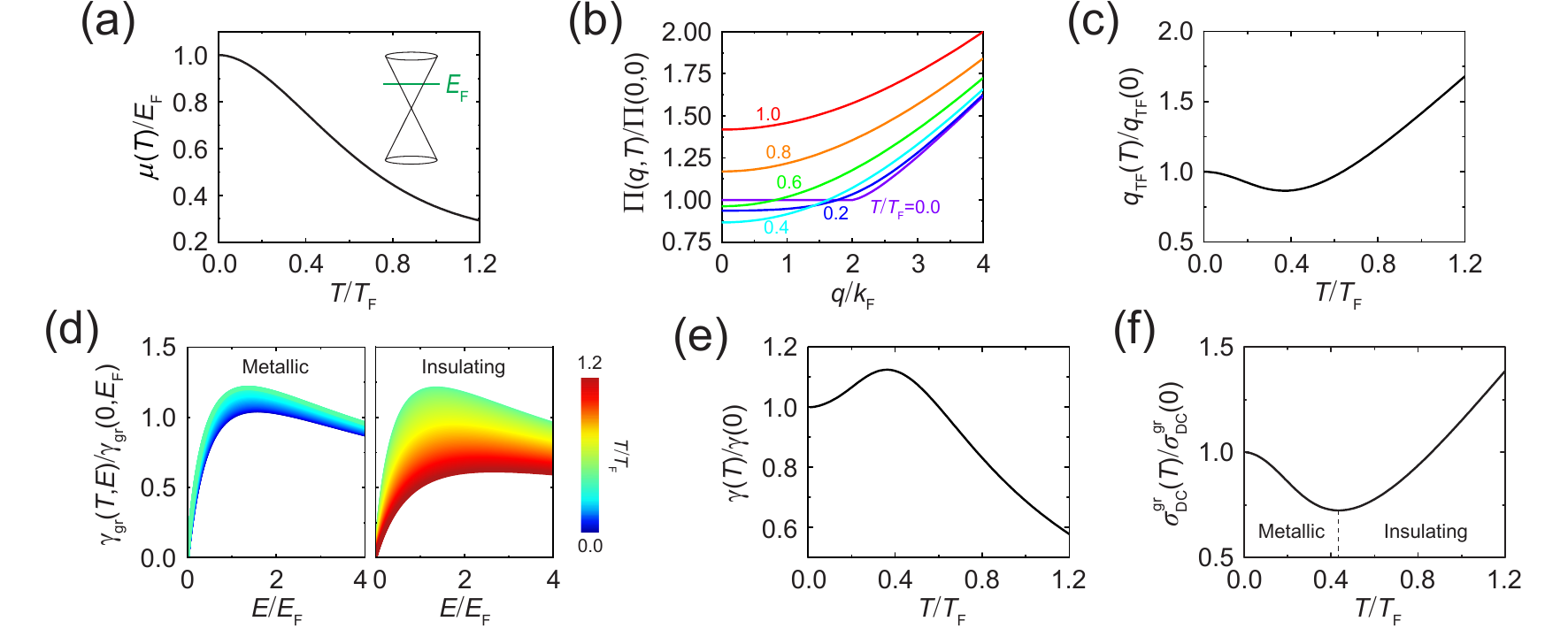} 
\caption{ 
\label{fig:graphene}
(a) Temperature-dependent chemical potential $\mu(T)/E_{\text{F}}$. 
Here, $E_{\text{F}}$ and $T_{\text{F}}$ denote Fermi energy and Fermi temperature, respectively.
(b) Calculated RPA polarizability function $\Pi(q,T)/\Pi(0,0)$ as a function of $q/k_{\text{F}}$ for several temperatures, which is consistent with previous report \cite{hwang_screening-induced_2009_2}. 
(c) Temperature-dependent Thomas-Fermi wave vector $q_{\text{TF}}(T)/q_{\text{TF}}(0)$. 
(d) Scattering rate of graphene $\gamma_{\text{gr}} (T,E)/\gamma_{\text{gr}}(0,E_{\text{F}})$ as a function of $E/E_{\text{F}}$ for the normalized temperature ($T/T_{\text{F}}$) range from 0.0 to 1.2, as indicated by the color bar. 
Here, we present $\gamma_{\text{gr}}(T,E)$ for two distinct regions, i.e., metallic and insulating regions.
(e),(f) Extracted temperature-dependent scattering rate $\gamma(T)/\gamma(0)$ 
and DC conductivity $\sigma_{\text{DC}}^{\text{gr}} (T)/\sigma_{\text{DC}}^{\text{gr}} (0)$ of high-mobility graphene. 
} 
\end{figure}

In this section, we discuss Dirac carrier dynamics of high-mobility graphene where the charged impurity scattering is the main scattering mechanism \cite{hwang_screening-induced_2009_2,bolotin_temperature-dependent_2008_2} and compare it with our Tb-doped system.
Based on Ref. \cite{hwang_screening-induced_2009_2}, we first calculate RPA polarizability function $\Pi(q,T)$ of graphene [Fig. \ref{fig:graphene}(b)].
At fixed non-zero temperature ($T/T_{\text{F}}\neq0$), $\Pi(q,T)$ increases as a function of wave vector $q$, which is similar to $\Pi(q,T)$ of our Tb-doped system in Fig. 4(g).
However, contrary to our Tb-doped system, $\Pi(q,T)$ of graphene shows a nonmonotonic $T$-depedence for $q < 2k_{\text{F}}$; there is a global minimum at $T/T_{\text{F}}\sim 0.4$,
which is well characterized by Thomas-Fermi wave vector $q_{\text{TF}}(T) \equiv \lim\limits_{q\to 0} qv_c\Pi(q,T)$ in Fig. \ref{fig:graphene}(c). 
With this $q_{\text{TF}}(T)$, we calculate the $T$-and $E$-dependent scattering rate $\gamma_{\text{gr}}(T,E)$ [Fig. \ref{fig:graphene}(d)];
it shows metallic behavior in the low-$T$ regime ($T/T_{\text{F}} \lesssim 0.4$), whereas it shows insulating behavior in the high-$T$ regime ($T/T_{\text{F}} \gtrsim 0.4$). 
Note that the low-$T$ ($T/T_{\text{F}} \sim 0.14$ at 100 K) insulating behavior of our Tb-doped system [Fig. 4(i)] corresponds to the high-$T$ insulating behavior of graphene. 
This difference arises from the existence of additional heavy hole band in our Tb-doped system compared with graphene of single Dirac band. 
Then, by calculating optical conductivity using $\gamma_{\text{gr}}(T,E)$ as input and fitting it to Drude model, we extract the temperature-dependent scattering rate $\gamma(T)$ and DC conductivity $\sigma_{\text{DC}}^{\text{gr}}(T)$ as shown in Fig. \ref{fig:graphene}(e) and \ref{fig:graphene}(f), respectively.
We find that both calculated $\gamma(T)$ and $\sigma_{\text{DC}}^{\text{gr}}(T)$ exhibit a nonmonotonic $T$-dependence, which well reflects the RPA polarizability function $\Pi(q,T)$ in Fig. \ref{fig:graphene}(b). 
As a result, we can understand the $T$-dependent Dirac carrier dynamics not only in the high-mobility graphene 
but also in our correlated Tb-doped iridate system via a single framework of $T$-dependent RPA screening effect of charged impurities.

\section{\normalsize TH\lowercase{z data and theoretical analysis} \lowercase{on the} L\lowercase{a-doped sample}}

\begin{figure}[h]
\includegraphics[width=\textwidth]{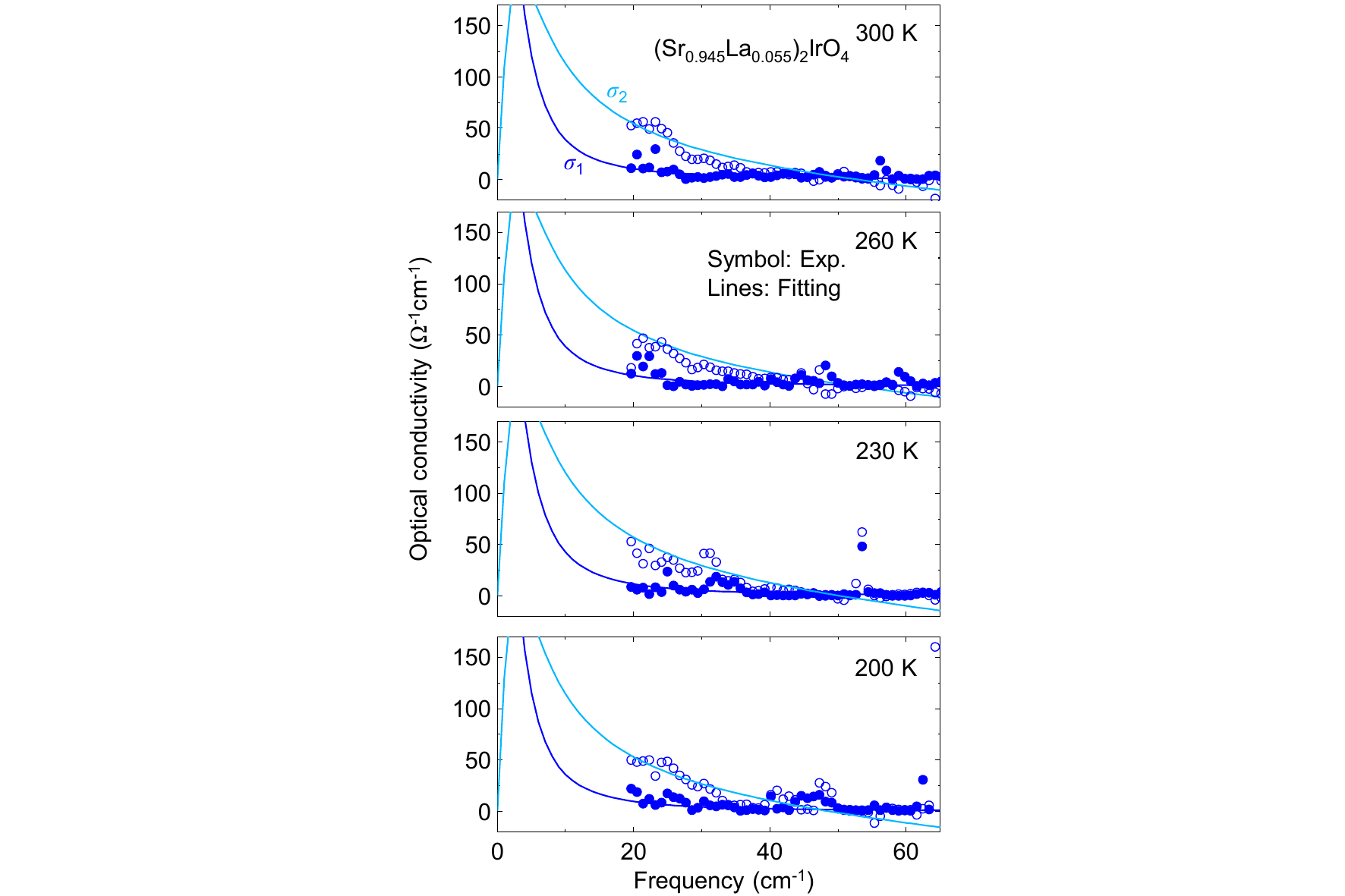} 
\caption{ 
\label{fig:THz_La}
Temperature-dependent complex optical conductivity spectra of (Sr$_{0.945}$La$_{0.055}$)$_2$IrO$_4$.
 Closed and open symbols indicate the real and imaginary parts of the optical conductivity, respectively. Lines denote the Drude fitting.} 
\end{figure}

In this section, we present THz data and theoretical analysis for the 5.5\% La-doped Sr$_2$IrO$_4$, i.e., (Sr$_{0.945}$La$_{0.055}$)$_2$IrO$_4$.
Because it has the same crystal structure and magnetic state as the Tb-doped system, the same symmetry analysis with d-wave order can be applied. 
Indeed, Dirac dispersion and a d-wave pseudogap were observed by ARPES for the 5\% La-doped system \cite{de_la_torre_collapse_2015_2}, suggesting that it has DPN ground state. 
Therefore, the $T$-dependent optical conductivity spectra of (Sr$_{0.945}$La$_{0.055}$)$_2$IrO$_4$ can be a good reference in discussing Dirac carrier dynamics in the correlated DSM in lightly-doped Sr$_2$IrO$_4$. 
Figure \ref{fig:THz_La} shows the $T$-dependent THz optical conductivity spectra of (Sr$_{0.945}$La$_{0.055}$)$_2$IrO$_4$. 
Real and imaginary parts of the optical conductivity are denoted by closed and open symbols, respectively. Drude fitting, indicated by solid lines, successfully reproduces the experimental data, suggesting that free carriers have a major contribution to the optical conductivity spectra in the THz region. 

\begin{figure}[h]
\includegraphics[width=\textwidth]{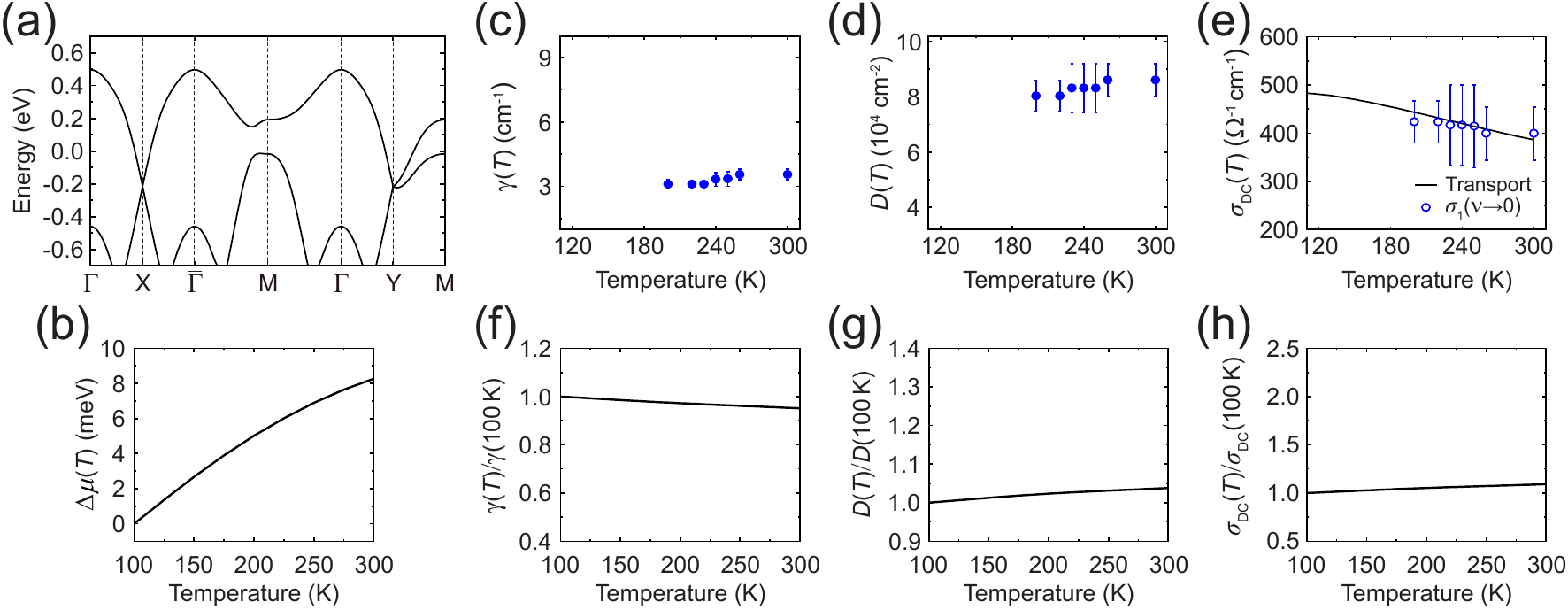} 
\caption{ 
\label{fig:eff_Ham_La}
Band structure and temperature-dependent electrodynamic parameters of (Sr$_{0.945}$La$_{0.055}$)$_2$IrO$_4$.
(a) Calculated band structure using $\Delta_d=30$ meV, which is consistent with Ref. \cite{zhou_correlation_2017_2}. 
(b) Calculated temperature-dependent chemical potential. 
(c),(d) Experimentally obtained scattering rate $\gamma(T)$ and Drude weight $D(T)$. 
These quantities are determined via the Drude-Lorentz fit of optical conductivity spectra at each $T$. 
(e) DC conductivity $\sigma_{\text{DC}}(T)$ estimated by Drude analysis (symbol) and directly measured by transport measurement (line). 
(f)-(h) Theoretically obtained $\gamma(T)$, $D(T)$, and $\sigma_{\text{DC}}(T)$, which are normalized by data at 100 K. } 
\end{figure}

Based on the parameters in Ref. \cite{zhou_correlation_2017_2}, we first calculate the band structure of DPN state to see the carrier dynamis of La-doped system [Fig. \ref{fig:eff_Ham_La}(a)]. 
The overall band structure is quite similar to that of Tb-doped system [Fig. 2(b)] and agrees well with the observed ARPES spectra \cite{de_la_torre_collapse_2015_2}. 
Compared with the Tb-doped system, a Dirac point is located at $E \sim -0.2$ eV and the top of parabolic band is more close to the Fermi level.
This causes significant difference in Dirac carrier dynamics between the two systems.
The chemical potential change of La-doped system is of about $8$ meV from 100 K to 300 K [Fig. \ref{fig:eff_Ham_La}(b)], which is three times smaller than that of Tb-doped system.
From these differences in $T$-dependences of the chemical potential as well as different energy of Dirac points, we expect that electrodynamic quantities are given differently beween the two systems.

By conducting a $T$-dependent THz experiment, we analyse the temperature dependence of the scattering rate $\gamma(T)$ and the Drude weight $D(T)$ in Figs. \ref{fig:eff_Ham_La}(c) and \ref{fig:eff_Ham_La}(d). 
The scattering rate $\gamma(T)$ appears to be very small, i.e., $3 ~ \text{cm}^{-1}$ in the entire $T$ range investigated, 
i.e., between 200 K and 300 K. 
When compared with the case of Sr$_2$(Ir$_{0.97}$Tb$_{0.03}$)O$_4$, $T$-dependence is relatively weaker, 
but the magnitude of $\gamma(T)$ is similarly small. 
Actually, as pointed out in the main text, we understand such a small $\gamma(T)$ as a hallmark of the Dirac fermion. Furthermore, the Drude weight $D(T)$ exhibits very similar $T$-dependent behaviors with the case of Sr$_2$(Ir$_{0.97}$Tb$_{0.03}$)O$_4$ [Fig. 4(b)]; 
it decreases as $T$ decreases. The combination between $\gamma(T)$ and $D(T)$ leads to the slightly decreasing behavior of zero-frequency value of the real part of optical conductivity $\sigma_1(\nu\rightarrow0)$ with increasing $T$, 
which is in good agreement with the transport data [Fig. \ref{fig:eff_Ham_La}(e)].

To understand THz data, we also calculate the optical conductivity for effective Hamiltonian using RPA scattering rate as we did for Tb-doped system.
Then we extract the temperature-dependent scattering rate $\gamma(T)$, Drude weight $D(T)$, and DC conductivity $\sigma_{\text{DC}}(T)$ by Drude fitting [Figs. \ref{fig:eff_Ham_La}(f)-\ref{fig:eff_Ham_La}(h)].
The extracted electrodynamic parameters $\gamma(T)$ and $D(T)$ are in good agreement with the experimental data. 
For $\sigma_{\text{DC}}(T)$, it also agrees well with $\sigma_1(\nu\rightarrow0)$ value within error bars.
Note that $\sigma_{\text{DC}}(T)$ is slightly different from transport data measured, which can be explained by possible phonon effects (we have checked numerically).
Compared to the Tb-doped system, the $T$-dependence of extracted electrodynamic parameters is very small, which is due to the smaller chemical potential change under $T$ variation and the deeper location of the Dirac point than Tb-doped system.
In summary, we can understand Dirac carrier dynamics of correlated DSM in lightly-doped Sr$_2$IrO$_4$, Sr$_2$(Ir$_{0.97}$Tb$_{0.03}$)O$_4$ and (Sr$_{0.945}$La$_{0.055}$)$_2$IrO$_4$, in a consistent way.

\end{document}